\documentclass[prl, reprint, longbibliography]{revtex4-1}
\usepackage{amsmath}
\usepackage{graphicx}
\usepackage{graphicx}
\usepackage{amsmath}
\usepackage{amssymb}
\usepackage{url}
\usepackage{bm}		
\usepackage{color}	
\usepackage{textcomp}
\usepackage[english]{babel}

\begin{document}
\newcommand{\drm}{\mathrm{d}}
\newcommand{\Lcal}{\mathcal{L}}
\newcommand{\eog}{\mathbf{e}_{og}}
\newcommand{\eogb}{\bar{\mathbf{e}}_{og}}
\newcommand{\eaj}{\mathbf{e}_{aj}}
\newcommand{\tb}[1]{\tilde{\mathbf{#1}}}


\title{Adjoint method and inverse design for nonlinear nanophotonic devices} 



\author{Tyler W. Hughes}
\thanks{These authors contributed equally.}

\author{Momchil Minkov}
\thanks{These authors contributed equally.}

\author{Ian A. D. Williamson}

\author{Shanhui Fan}
\email[]{shanhui@stanford.edu}
\affiliation{Department of Electrical Engineering, and Ginzton Laboratory, Stanford University, Stanford, CA 94305, USA}

\date{\today}

\begin{abstract}
The development of inverse design, where computational optimization techniques are used to design devices based on certain specifications, has led to the discovery of many compact, non-intuitive structures with superior performance. Among various methods, large-scale, gradient-based optimization techniques have been one of the most important ways to design a structure containing a vast number of degrees of freedom. These techniques are made possible by the adjoint method, in which the gradient of an objective function with respect to all design degrees of freedom can be computed using only two full-field simulations. However, this approach has so far mostly been applied to linear photonic devices. Here, we present an extension of this method to modeling nonlinear devices in the frequency domain, with the nonlinear response directly included in the gradient computation. As illustrations, we use the method to devise compact photonic switches in a Kerr nonlinear material, in which low-power and high-power pulses are routed in different directions. Our technique may lead to the development of novel compact nonlinear photonic devices.
\end{abstract}

\maketitle

In recent years, there has been significant interest in using computational optimization tools to design novel nanophotonic devices with a wide range of applications \cite{lalau-keraly_adjoint_2013,wang_adjoint-based_2018,elesin_design_2012,piggott_fabrication-constrained_2017,piggott_inverse_2015,kao_maximizing_2005,hughes_method_2017, molesky_outlook_2018,sigmund_systematic_2003,matzen_systematic_2011,jensen_topology_2005,frellsen_topology_2016, shen_integrated-nanophotonics_2015,shen_design_2003, Minkov2014, Minkov2015, shi_optimization_2018,veronis_method_2004}. Much of this progress \cite{lalau-keraly_adjoint_2013,wang_adjoint-based_2018,elesin_design_2012,piggott_fabrication-constrained_2017,piggott_inverse_2015,kao_maximizing_2005,hughes_method_2017, molesky_outlook_2018,sigmund_systematic_2003,matzen_systematic_2011,jensen_topology_2005,frellsen_topology_2016} is made possible by the adjoint method \cite{giles2000introduction, veronis_method_2004}, a technique which allows the gradient of an objective function to be computed with respect to an arbitrarily large number of degrees of freedom using only two full-field simulations.  This method
makes large-scale gradient-based design of electromagnetic structures possible. When
compared to brute force searching through the parameter space \cite{shen_integrated-nanophotonics_2015}, and other commonly used design methods,
like stochastic global optimization algorithms \cite{shen_design_2003, Minkov2014, Minkov2015, shi_optimization_2018}, gradient-based design has a number of practical advantages. For example, a very large number of design parameters can be adjusted simultaneously, and the number of structures one is required to evaluate in order to reach a high-performing structure can be far smaller compared with the total number of structures in the search space. 

Up to now, in photonics the adjoint method has been mostly applied to gradient-based optimization of linear optical devices. The generalization of the adjoint method to nonlinear optical devices would create new possibilities in several exciting fields such as on-chip lasers \cite{yamashita_raman_2015}, frequency combs \cite{okawachi_octave-spanning_2011}, spectroscopy \cite{moon_absolute_1997}, neural computing \cite{khoram2018stochastic}, and quantum information processing \cite{guo_-chip_2016}.  To this end, several recent works \cite{lin_cavity-enhanced_2016,lin_topology_2017,bravo-abad_enhanced_2007} have applied adjoint methods to engineer \textit{linear} devices to display favorable properties for nonlinear optical applications, such as high quality factors, small mode volume, or large field overlap between the modes of interest.  However, these works do not directly optimize the nonlinear systems.

To solve for the adjoint sensitivity of a nonlinear system, the standard option is to work within a time-domain adjoint formalism, which entails simulating an additional linear system with a time-varying permittivity \cite{elesin_design_2012}.  However, as this formalism requires the storage of the fields at each time step, it has substantial memory requirements.  Furthermore, because in many cases the steady-state behavior of the system is of interest, a frequency-domain approach is preferred as the steady state response can be obtained directly, without the need for going through a large number of time steps as in a time-domain simulation.  The general mathematical formalism for the adjoint method in nonlinear systems is known in the applied mathematics literature \cite{Strang2007}. But, with the exception of a very recent preprint that seeks to design a nonlinear element in an optical neural network  \cite{khoram2018stochastic}, such a formalism has not been previously applied to nonlinear photonic device optimizations.

In this work we outline, in detail, how the adjoint method may be used to optimize the steady-state response of a nonlinear optical device in the frequency domain. We first outline a formalism for generalizing adjoint problems to arbitrary nonlinear problems.  Then, as a demonstration, we use our method to inverse-design photonic switches with Kerr nonlinearity.  Our results may be applied more generally to other objective functions and sources of nonlinearity and provides new possibilities for designing novel nonlinear optical devices.

\section{\label{sec:nl_avm}Nonlinear adjoint method}

We first outline the formulation of the adjoint method for the inverse design of nonlinear optical devices.  The goal of inverse design is to find a set of real-valued design variables $\bm{\varphi}$ that maximize a real-valued objective function $\mathbf{\Lcal} = \mathbf{\Lcal}(\mathbf{e}, \mathbf{e}^*, \bm{\varphi})$, where the complex-valued vector $\mathbf{e}$ is given by the solution to the equation
\begin{align}
\mathbf{f}(\mathbf{e}, \mathbf{e}^*, \bm{\varphi}) &= 0
\label{eq:constraints}.
\end{align}
For example, eq. (\ref{eq:constraints}) may represent the steady-state Maxwell's equations with an intensity-dependent permittivity distribution where $\mathbf{e}$ is the electric field distribution.  The solution to eq. (\ref{eq:constraints}) may be found with any nonlinear equation solver, such as with the Newton-Raphson method \cite{press2007numerical}.  We further note that the treatment of $\mathbf{e}$ and its complex conjugate as independent variables is necessary for differentiation as will be shown later.

The aim of the optimization is to maximize the objective function with respect to the design variables $\bm{\varphi}$. For this purpose, it is essential to compute the sensitivity of $\Lcal$ with respect to each element of $\bm{\varphi}$.  For simplicity, we derive the derivative of the objective function with respect to a single parameter $\varphi$, which is written
\begin{equation}
\frac{d\mathcal{L}}{d\varphi} = \frac{\partial\mathcal{L}}{\partial\varphi} + \frac{\partial\mathcal{L}}{\partial\mathbf{e}}\frac{d\mathbf{e}}{d\varphi} + \frac{\partial\mathcal{L}}{\partial\mathbf{e}^*}\frac{d\mathbf{e}^*}{d\varphi}.
\end{equation}
Or, in matrix form as
\begin{equation}
\frac{d\mathcal{L}}{d\varphi} = \frac{\partial\mathcal{L}}{\partial\varphi} + 
\begin{bmatrix}
\partial\mathcal{L} / \partial\mathbf{e} &
\partial\mathcal{L} / \partial\mathbf{e}^*
\end{bmatrix}
\begin{bmatrix}
d\mathbf{e} / d\varphi \\ d\mathbf{e}^* / d\varphi
\end{bmatrix}.
\label{eq:dLdphi_matrix}
\end{equation}
To compute $d\mathbf{e}/d\varphi$ and $d\mathbf{e}^*/d\varphi$, we differentiate eq. (\ref{eq:constraints}):
\begin{align}
\frac{d\mathbf{f}}{d\varphi} &= 0 = \frac{\partial \mathbf{f}}{\partial \varphi} + \frac{\partial \mathbf{f}}{\partial \mathbf{e}}\frac{d \mathbf{e}}{d\varphi} + \frac{\partial \mathbf{f}}{\partial \mathbf{e}^*}\frac{d \mathbf{e}^*}{d\varphi}. \label{eq:dfdphi}
\end{align}
Eq. (\ref{eq:dfdphi}) together with its complex conjugate then yields
\begin{equation}
\begin{bmatrix}
\partial \mathbf{f} / \partial \mathbf{e} &
\partial \mathbf{f} / \partial \mathbf{e}^* \\
\partial \mathbf{f}^* / \partial \mathbf{e} &
\partial \mathbf{f}^* / \partial \mathbf{e}^*
\end{bmatrix}
\begin{bmatrix}
d\mathbf{e} / d\varphi \\ d\mathbf{e}^* / d\varphi
\end{bmatrix} = -\begin{bmatrix}
\partial \mathbf{f} / \partial \varphi \\ \partial \mathbf{f}^* / \partial \varphi
\end{bmatrix}.
\end{equation}
Thus, formally we can rewrite eq. (\ref{eq:dLdphi_matrix}) as
\begin{align}
& \frac{d\mathcal{L}}{d\varphi} = \frac{\partial\mathcal{L}}{\partial\varphi} - \label{eq:dLdphi}\\
& \begin{bmatrix}
\partial\mathcal{L} / \partial\mathbf{e} &
\partial\mathcal{L} / \partial\mathbf{e}^*
\end{bmatrix}
\begin{bmatrix}
\partial \mathbf{f} / \partial \mathbf{e} &
\partial \mathbf{f} / \partial \mathbf{e}^* \\
\partial \mathbf{f}^* / \partial \mathbf{e} &
\partial \mathbf{f}^* / \partial \mathbf{e}^*
\end{bmatrix}^{-1}
\begin{bmatrix}
\partial \mathbf{f} / \partial \varphi \\ \partial \mathbf{f}^* / \partial \varphi
\end{bmatrix}.
\nonumber
\end{align}

In analogy with the linear adjoint method, we can now compute the gradient by solving an additional linear system.  We define a complex-valued adjoint field $\eaj$ as the solution to
\begin{align}
\begin{bmatrix}
\partial \mathbf{f} / \partial \mathbf{e} &
\partial \mathbf{f} / \partial \mathbf{e}^* \\
\partial \mathbf{f}^* / \partial \mathbf{e} &
\partial \mathbf{f}^* / \partial \mathbf{e}^*
\end{bmatrix}^{T}
\begin{bmatrix}
\eaj \\ \eaj^*
\end{bmatrix} = - \begin{bmatrix}
{\partial\mathcal{L} / \partial\mathbf{e}}^T \\
{\partial\mathcal{L} / \partial\mathbf{e}^*}^T
\end{bmatrix}, 
\label{eq:eaj}
\end{align}
and the gradient of the objective function is then
\begin{align}
\frac{d\mathcal{L}}{d\varphi} = \frac{\partial\mathcal{L}}{\partial\varphi} + 2\mathcal{R}\left(
\eaj^T
\frac{\partial \mathbf{f}} { \partial \varphi} \right)
\label{eq:dLdphi_eaj},
\end{align}
where $\mathcal{R}$ denotes taking the real part. In deriving eq. (\ref{eq:dLdphi_eaj}), we have used the fact that both $\Lcal$ and $\varphi$ are real.  In the case of multiple parameters $\bm{\varphi}$, we can simply replace $\partial \mathbf{f} / \partial \varphi$ with the matrix $\partial \mathbf{f} / \partial \bm{\varphi}$. Since $\eaj$ only needs to be solved once regardless of the number of parameters, gradients may be computed with very little marginal cost for an arbitrary number of free parameters, making large-scale, gradient-based optimization possible.

\section{Application to Kerr Nonlinearity}

\begin{figure}[t]
\includegraphics[width=\columnwidth]{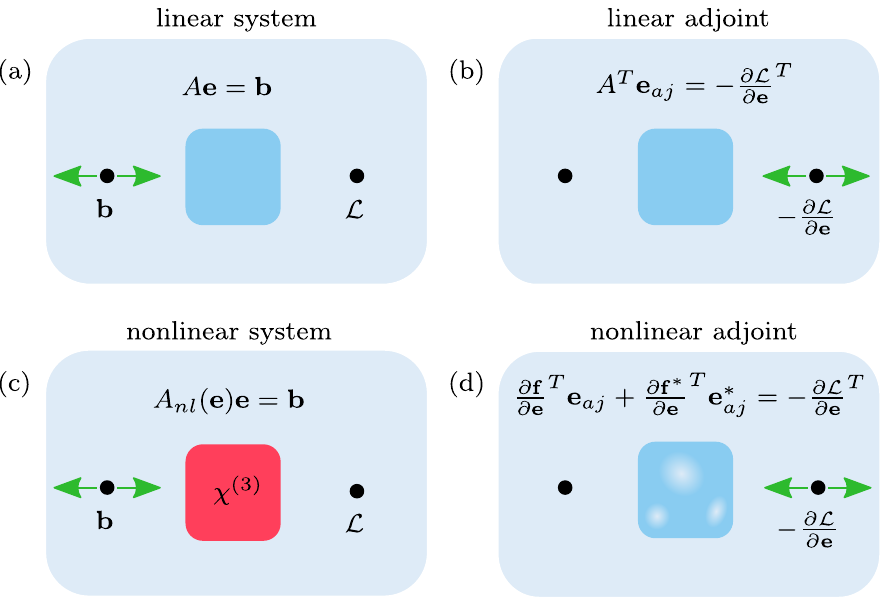}
\caption{\label{fig:avm} Illustration of the adjoint field computation for a linear and a nonlinear system. (a) The linear system driven by a point source $\mathbf{b}$ with an objective function $\mathcal{L}$  given by the field intensity at a measuring point. (b)  The adjoint problem for the linear system: the same system driven by a point source given by $-\partial\mathcal{L}/\partial \mathbf{e}$ located at the measuring point.  (c) The nonlinear system containing a medium with Kerr nonlinearity (red).  The electric fields are the solution to a nonlinear equation. (d) The adjoint problem for the nonlinear system, which is a \textit{linear} system of equations for the adjoint field and its complex conjugate. The Kerr medium is replaced by a linear region whose permittivity depends on the nonlinear fields.}
\end{figure}

We now apply the general formalism as discussed above to the  optimization of nonlinear optical systems. Since the formalism is applicable to linear optical systems as well, for illustration purposes here we use it to treat both the linear and the nonlinear cases, in order to highlight aspects that are unique to nonlinear systems.   A schematic outlining the two cases is presented in Fig. \ref{fig:avm}.  For a linear system, Maxwell's equations for the steady-state at a frequency $\omega_0$ may be written as 
\begin{align}
\mu_0^{-1}  \nabla &\times \nabla \times \mathbf{E}(\mathbf{r}) - \omega_0^2 \epsilon_0 \epsilon_r(\mathbf{r}) \mathbf{E}(\mathbf{r})  = -i \omega_0 \mathbf{J}(\mathbf{r}),
\label{eq:fdfd_lin}
\end{align}
where $\mathbf{E}(\mathbf{r})$ is the electric field, $\mathbf{J}(\mathbf{r})$ is the electric current source, $\epsilon_r(\mathbf{r})$ is the relative dielectric permittivity, and we have assumed relative permeability $\mu_r = 1$ everywhere. Compactly, and to make connection to the general formalism in the previous section, this can be written in matrix form as
\begin{equation}
\mathbf{f}(\mathbf{e}, \mathbf{e}^*, \bm{\varphi}) = A(\bm{\epsilon}_r) \mathbf{e} - \mathbf{b} = 0,
\label{eq:lin_forward}
\end{equation}
where $A$ is a linear operator, vectors $\mathbf{e}$ and $\bm{\epsilon}_r$ now contain the electric fields and the relative permittivity, respectively, and $\mathbf{b}$ is a vector proportional to the current source. The design parameters $\bm{\varphi}$ in this case is the permittivity distribution $\bm{\epsilon}_r$. Eq. (\ref{eq:lin_forward}) can be solved to obtain the electric fields $\mathbf{e}$, as diagrammed by Fig. \ref{fig:avm}(a).

We assume an objective function $\mathcal{L}$ that depends on the field solution to eq. (\ref{eq:lin_forward}) and we take the linear relative permittivity distribution as the set of design variables.  Because $\partial \mathbf{f} / \partial \mathbf{e} = A$ and $\partial \mathbf{f} / \partial \mathbf{e}^* = 0$ for the linear system, from eq. (\ref{eq:eaj}), the adjoint field may be written simply as the solution to the equation
\begin{equation}
A^T(\bm{\epsilon}_r) \eaj = -\left( \partial \mathcal{L}/\partial \mathbf{e}\right)^T,
\end{equation}
as shown in Fig. \ref{fig:avm}(b).  For a reciprocal system, $A^T = A$, thus the original and the adjoint fields are solutions to the same linear problem but with different source terms. Note that the source for the adjoint field, $-\left( \partial \mathcal{L}/\partial \mathbf{e}\right)^T$ depends on both the objective function and the original solution. 

Once the adjoint field is computed, the gradient of the objective function with respect to the permittivity distribution is given, through eq. (\ref{eq:dLdphi_eaj}), by 
\begin{align}
\frac{d\mathcal{L}}{d\bm{\epsilon}_r} &= \frac{\partial\mathcal{L}}{\partial \bm{\epsilon}_r} + 2\mathcal{R}\left(
\eaj^T
\frac{\partial A} { \partial \bm{\epsilon}_r} \mathbf{e} \right) \\
 &= \frac{\partial\mathcal{L}}{\partial \bm{\epsilon}_r} - 2 \omega_0^2 \epsilon_0 \mathcal{R}\left(
\eaj^T
 \mathbf{e} \right).
\label{eq:dLdphi_lin}
\end{align}

Having reviewed the adjoint formalism for linear optical systems we now consider nonlinear optical systems. As an example, we introduce Kerr nonlinearity into the system \cite{Boyd__2008}, which corresponds to an intensity-dependent permittivity
\begin{equation}
\tilde{\epsilon}_r(\mathbf{r}) = \epsilon_r(\mathbf{r}) + 3\omega_0^2 \epsilon_0\chi^{(3)}(\mathbf{r}) \left|\mathbf{E}(\mathbf{r}) \right|^2,
\label{eq:eps_mod}
\end{equation}
where $\chi^{(3)}(\mathbf{r})$ is the nonlinear susceptibility distribution. Other types of nonlinear terms can also be treated with the formalism outlined above. Replacing $\epsilon_r(\mathbf{r})$ in Eq. (\ref{eq:fdfd_lin}) with $\tilde{\epsilon}_r(\mathbf{r})$ in Eq. (\ref{eq:eps_mod}), our system is then described by the equation:
\begin{equation}
\mathbf{f}(\mathbf{e}, \mathbf{e}^*, \bm{\varphi}) = A_{nl}(\bm{\epsilon}_r, \bm{\chi}, \mathbf{e}) \mathbf{e} - \mathbf{b} = 0,
\label{eq:nl_forward}
\end{equation}
where $A_{nl} \equiv \left[ A(\bm{\epsilon}_r) - \textrm{diag}\left( \bm{\chi} \odot |
\mathbf{e}|^2 \right) \right]$.  Here, $\odot$ is element-wise vector multiplication and $\textrm{diag}(\mathbf{v})$ represents a diagonal matrix with vector $\mathbf{v}$ on the main diagonal.  The vector $\bm{\chi}$ corresponds to the term $3\omega_0^2\epsilon_0 \chi^{(3)}(\mathbf{r})$ and
$|\mathbf{e}|^2 \equiv \mathbf{e} \odot \mathbf{e}^*$.  Again, for concreteness, the design parameters $\bm{\varphi}$ correspond to the permittivity $\bm{\epsilon}_r$. The solution to this problem is diagrammed in Fig. \ref{fig:avm}(c).

From eq. (\ref{eq:eaj}) we may now compute the partial derivatives of $\mathbf{f}$ with respect to the electric fields $\mathbf{e}$, which is needed to construct the adjoint problem.
\begin{align}
\partial \mathbf{f} / \partial \mathbf{e} &= A - 2 \textrm{diag}\left( \bm{\chi} \odot |\mathbf{e}|^2  \right)\\
\partial \mathbf{f}^* / \partial \mathbf{e} &= -\textrm{diag}\left( \bm{\chi} \odot \mathbf{e}^* \odot \mathbf{e}^* \right).
\end{align}
With this, we then express the adjoint field as a solution to the linear system
\begin{equation}
\left( \partial \mathbf{f} / \partial \mathbf{e} \right)^T \mathbf{e}_{aj} + \left( \partial \mathbf{f}^* / \partial \mathbf{e} \right)^T \mathbf{e}_{aj}^* = -\left( \partial \Lcal/\partial \mathbf{e}\right)^T,
\label{eq:adjoint_nonlin}
\end{equation}
which is diagrammed in Fig. \ref{fig:avm}(d).

For a nonlinear system, to obtain the field $\mathbf{e}$, one will need to solve a nonlinear equation (e.g. eq. (\ref{eq:nl_forward})). However, we emphasize that the adjoint problem, as required to determine the derivative of the objective function, is a \textit{linear} problem. The size of the adjoint problem is twice as large as the corresponding linear problem of Eq. (\ref{eq:lin_forward}), but it is of a similar form, with the source dependent upon the solution $\mathbf{e}$.   

Once the adjoint field is computed, the gradient $d\Lcal/d\bm{\epsilon}_r$ is evaluated from eq. (\ref{eq:dLdphi_lin}) as in the linear case.  Here for simplicity we do not assume any explicit dependence of the nonlinearity on the design variable. However, the formalism is straightforward to extend to that case, as explained in the Supplementary Information.

\section{Inverse Design of Optical Switches}

\begin{figure*}[t]
\centering
\includegraphics[width=0.8\textwidth]{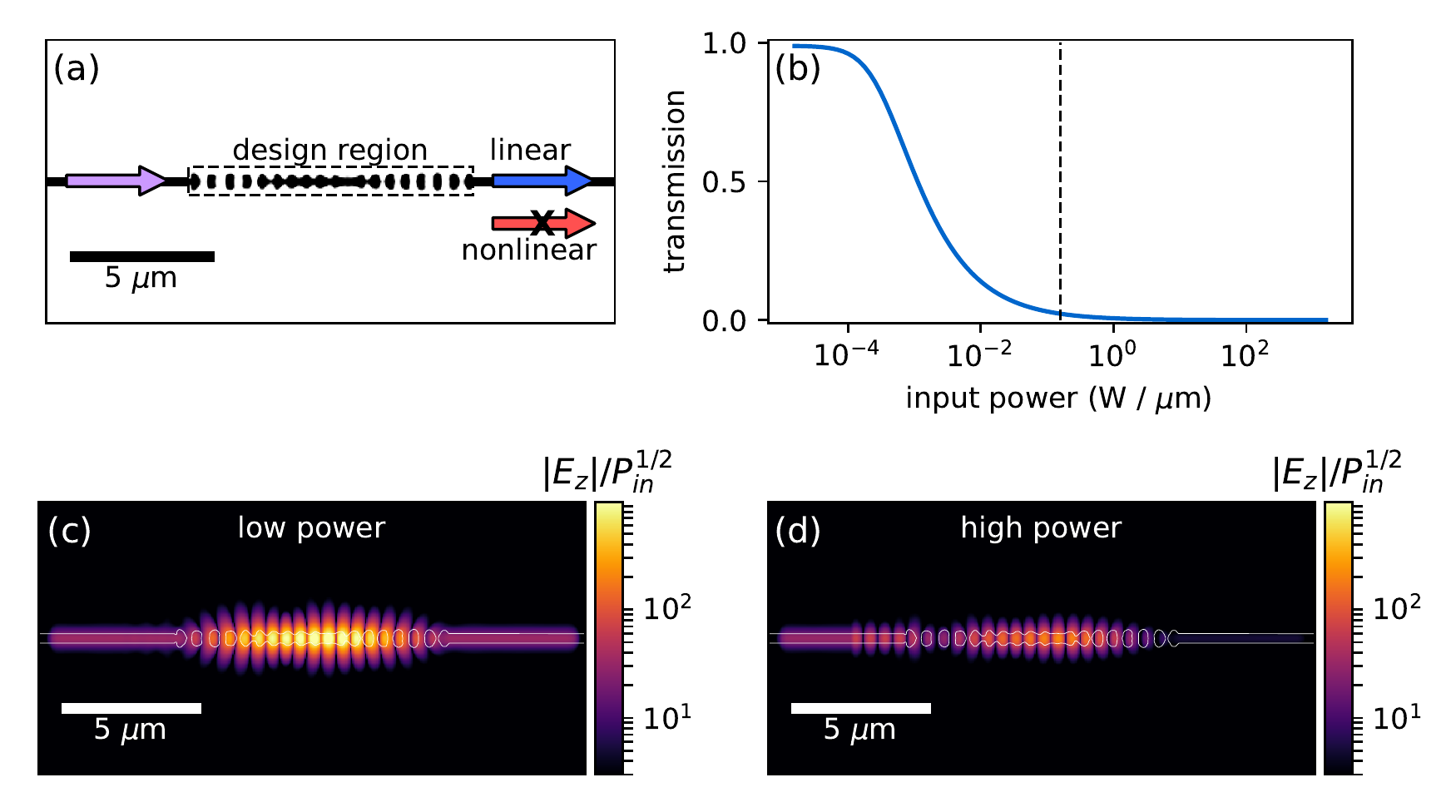}
\caption{\label{fig:2_port} \textbf{Inverse design demonstration of a $1 \to 1$ port switch.}  (a) Optical power is input into the left port (purple arrow). The goal of optimizing the design region (blue square) is to maximize power transmission in the linear regime (blue arrow) and minimize transmission in the nonlinear regime (red arrow).  The final permittivity distribution after optimizing is also shown. The black regions are chalcogenide with a relative permittivity of 5.95 and a $\chi^{(3)}$ of $4.1\times 10^{-19}$ m$^2/$V$^{2}$. The waveguide regions outside the design region have a width of $0.3 \mu$m. The operating wavelength is $2\mu$m. (b) The transmission as a function of input power, demonstrating the switching behavior at around $10^{-3}$ W/$\mu$m.  The dashed black line indicates the input power used for the high-power regime in the optimization. (c-d) The amplitude of the simulated electric field of the final structure, in the linear (c) and nonlinear (d) regimes, respectively -- with input power of $10^{-9}$ W/$\mu$m and $0.157$ W/$\mu$m, respectively. $E_z$ corresponds to the out-of-plane electric field in the 2D simulation.}
\end{figure*}

\begin{figure*}[t]
\centering
\includegraphics[width=0.8\textwidth]{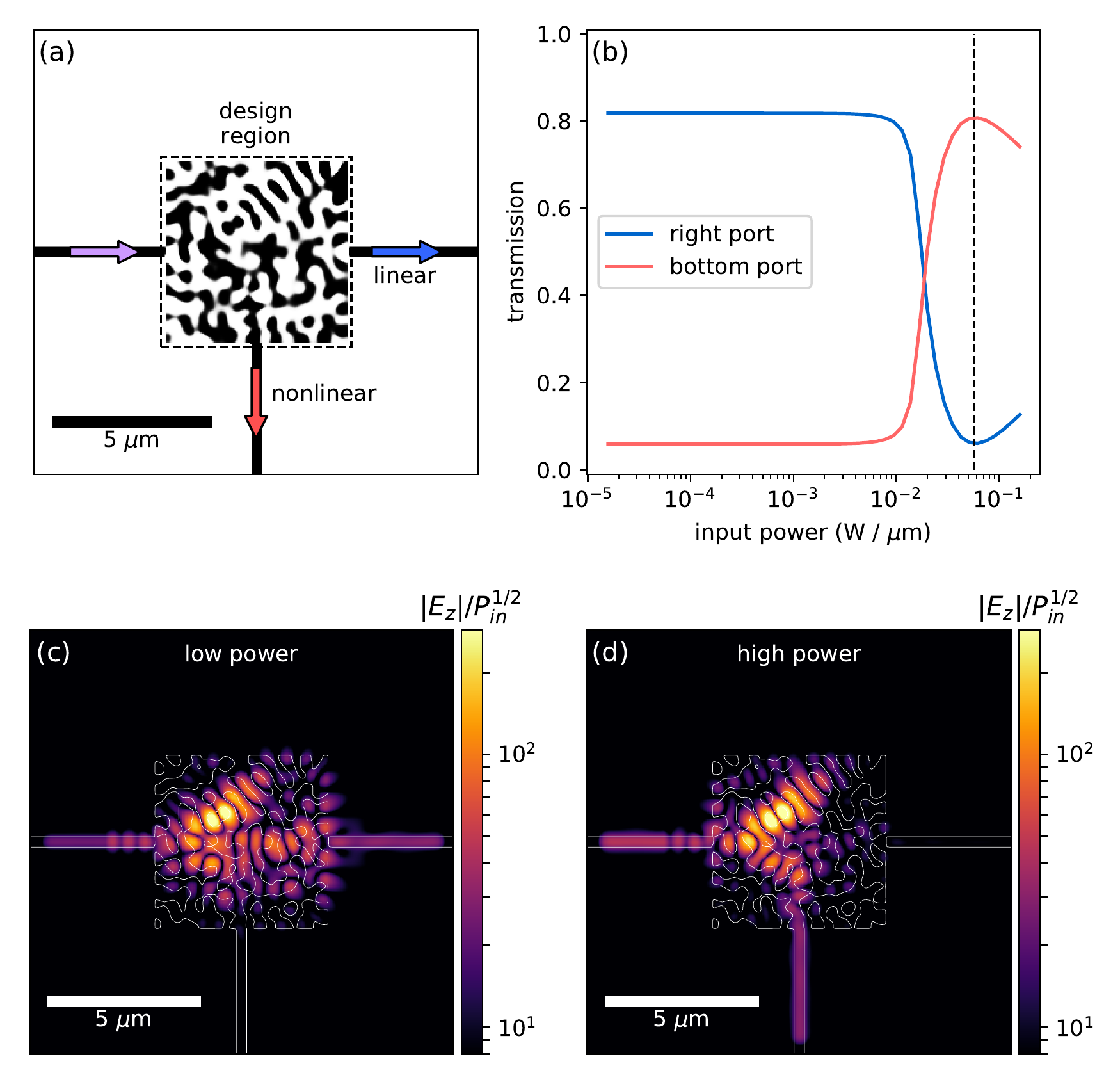}
\caption{\label{fig:t_port} \textbf{Inverse design demonstration of a $1 \to 2$ port switch.}  (a) Optical power is input into the left port (purple arrow). The goal of optimizing the design region (blue square) is to maximize the power transmission to the right port (blue arrow) in the linear regime and maximize transmission to the bottom port (red arrow) in the nonlinear regime. The final permittivity distribution after optimizing is also shown. (b) The transmission in the right (blue) and bottom (red) ports as a function of input power, demonstrating the switching behavior at around $2\times10^{-2}$ W/$\mu$m.  The dashed black line indicates the input power used for the high-power regime in the optimization. (c-d) The amplitude of the simulated electric field of the final structure, in the linear (c) and nonlinear (d) regimes, respectively -- with input power of $10^{-9}$ W/$\mu$m and $0.057$ W/$\mu$m, respectively.}
\end{figure*}

We now demonstrate the use of this nonlinear adjoint formalism to inverse design optical switches with desired power-dependent performance characteristics.  In Figs. \ref{fig:2_port} and \ref{fig:t_port}, we show the optimization procedures and performance characteristics of a 1 $\to$ 1 and 1 $\to$ 2 port device, respectively. The operating frequency for both devices correspond to a free-space wavelength of 2$\mu$m. 

For each device, we seek to maximize the corresponding objective function with respect to the permittivity distribution within a fixed design region.  To perform the numerical optimization of the structure, we use the finite-difference frequency-domain method (FDFD) \cite{shin2012choice}, where the fields and operators of Eq. (\ref{eq:fdfd_lin}) are spatially discretized using a Yee lattice \cite{yee1966numerical}. For simplicity, we restrict our study to two-dimensional structures (i.e. structures with infinite extent in the third dimension), and transverse-magnetic polarization, which has only non-zero out-of-plane electric field components. In the optimization process, we start with an initial relative permittivity in the design region.  We solve the electric field distribution in the structure by solving the nonlinear equation (eq. (\ref{eq:nl_forward})).  Then, we compute the gradient of $\mathcal{L}$ with respect to the relative permittivity distribution in this region using eq. (\ref{eq:dLdphi_lin}).  With the gradient information, we perform updates of the design variables using the limited-memory BFGS \cite{byrd1995limited} algorithm, although a simple gradient ascent algorithm would also suffice.  This procedure is repeated until convergence on a final structure. 

We choose optimization parameters corresponding to a device made from chalcogenide glass (Al$_2$S$_3$), which exhibits a strong $\chi^{(3)}$ response and high damage threshold \cite{White_OptLett_2011,Lamont_OptExpress_2008,Boyd__2008}.  During the optimization, the relative permittivity was constrained to lie between $1$ (air) and $5.95$ (Al$_2$S$_3$).  We further assume that the materials exhibit nonlinearity only within the design regions outlined in Fig. \ref{fig:2_port}(a) and \ref{fig:t_port}(a).

To create a more realistic final structure, the strength of the nonlinear susceptibility was assumed to be proportional to the "density" of material, $\rho$, defined as 
\begin{equation}
\rho(\mathbf{r})= \frac{\epsilon_r(\mathbf{r}) - 1}{\epsilon_m - 1 },
\label{eq:rho_def}
\end{equation}
where $\epsilon_m$ is the permittivity of the material. This assumption ensures that air regions do not exhibit a nonlinear refractive index. Eq. (\ref{eq:rho_def}) adds an $\epsilon_r$ dependence in the nonlinear susceptibility, which is straightforwardly treated in the adjoint method, as discussed in the Supplementary Information.  Low-pass spatial filtering and projection techniques \cite{zhou2015minimum} were applied during optimization to create binarized (air and chalcogenide) final structures with large, smoothed features.  Additional details on this are described in the Supplementary Information.

Our first device, as shown in Fig. \ref{fig:2_port}, consists of a waveguide-fed 1 $\to$ 1 port geometry with a central design region.  We optimize this design region to maximize power transmission in the linear regime when the incident power is sufficiently low such that the nonlinear terms do not affect the transmission, and minimize transmission in the nonlinear regime when the incident power is at a specific high value such that the nonlinearity plays a significant role.  This corresponds to an objective function of the form
\begin{equation}
    \Lcal(\mathbf{e}_{\textrm{low}}, \mathbf{e}_{\textrm{high}}) = \left|\mathbf{m}^T \mathbf{e}_{\textrm{low}}\right|^2 - \left|\mathbf{m}^T \mathbf{e}_{\textrm{high}}\right|^2,
\end{equation}
where $\mathbf{e}_{\mathrm{low}}$ and $\mathbf{e}_{\mathrm{high}}$ are the simulated fields with a low and a high input power, respectively, $\mathbf{m}$ is the modal profile of the electric field for the waveguide in the output port, and the objective function is normalized such that its maximum value is 1.  The optimization setup and the optimized structure are diagrammed in Fig. \ref{fig:2_port}(a). The final structure resembles a resonator between two Bragg mirrors, effectively acting like a bistable switch \cite{soljacic2002optimal, yanik_all-optical_2003}. Fig. \ref{fig:2_port}(b) shows the transmission as a function of the input power, and it clearly switches from high to low as the input power increases. This is also illustrated in panels (c)-(d), where we plot the field amplitude distributions in the low-power (high-transmission) regime and in the high-power (low-transmission) regime, respectively. The computed power transmission coefficients for these two panels are 98.2\% and 3.1\%, respectively. The value of the input power used in the optimization and in panel (d) is shown by a dashed line in panel (b). At this input power, the device exhibits a maximum nonlinear refractive index shift of $4.0 \times 10^{-3}$, which is below the damage threshold for Al$_2$S$_3$ using sub-nanosecond pulses \cite{chorel2018robust} (see Supplementary Information). The transmission spectrum of this structure gives a resonance peak with a full-width at half maximum of 38GHz (see Supplementary Information). In the Supplementary Information, we also list the specific optimization parameters, and show the value of the objective function during the optimization process. Reaching the final optimized structure shown in Fig. \ref{fig:2_port} required the evaluation of 2000 structures, but a reasonably high-performing structure is already reached after only a few hundred iterations.  

We also use the same technique for the inverse design of a 1 $\to$ 2 port switch where light is guided to the right port in the linear regime and to the bottom port in the nonlinear regime.  To achieve this design, we define the objective function as
\begin{align}
    \Lcal(\mathbf{e}_\textrm{low}, \mathbf{e}_\textrm{high}) =~ & \left|\mathbf{m}_\textrm{r}^T \mathbf{e}_\textrm{low}\right|^2 - \left|\mathbf{m}_\textrm{r}^T \mathbf{e}_\textrm{high}\right|^2 \\ \nonumber
    - &\left|\mathbf{m}_\textrm{b}^T \mathbf{e}_\textrm{low}\right|^2 + \left|\mathbf{m}_\textrm{b}^T \mathbf{e}_\textrm{high}\right|^2 ,
\end{align}
where $\mathbf{m}_\textrm{r}$ and $\mathbf{m}_\textrm{b}$ denote the mode profiles of the waveguides in the right and bottom ports, respectively. These are normalized such that the objective function has a maximum value of 1 for a perfect switching operation.  The setup of the optimization problem and the final design are diagrammed in Fig. \ref{fig:t_port}(a). We note that the device displays a non-intuitive geometry while retaining large features and good binarization.

In Fig. \ref{fig:t_port}(b), we plot the transmission through the right and through the bottom ports as a function of input power. This clearly shows the switching of power from the right port to the bottom port in the linear and nonlinear regimes, respectively.  Specifically, in the linear regime, the device has a power transmission of 81.8\% and 5.9\% to the right and bottom ports, respectively, while in the nonlinear regime, at the input power marked by the dashed line in Fig. \ref{fig:t_port}(b), these values are 6.1\% and 80.8\%, respectively. The electric field amplitudes for linear and nonlinear regimes are displayed in \ref{fig:t_port}(c)-(d). The operational bandwidth for this device is approximately 90 GHz (see Supplementary Information).  The device exhibits a maximum nonlinear refractive index shift of $3.9 \times 10^{-3}$, which is also below the acceptable damage threshold for Al$_2$S$_3$ using sub-nanosecond pulses \cite{chorel2018robust}.  A full list of optimization parameters and a plot of the objective function vs. iteration number is shown in the Supplementary Information.

\section{Discussion}

We have presented an extension to the adjoint variable method applied to the optimization of an electromagnetic system with Kerr nonlinearity. Our approach can be straightforwardly applied to other types of nonlinearities which do not mix frequencies, such as saturable gain or absorption. Moreover, the methods here should be straightforwardly generalizable to treat nonlinear problems involving frequency mixing. For example, one can imagine implementing a similar adjoint method in combination with the multi-frequency finite-difference frequency-domain implementations for nonlinear wave interactions \cite{Shi2016multi}. 

In addition to the design of optical switches, our formalism may prove useful for many other interesting problems in nonlinear photonics.  For example, one could apply our approach to design nonlinear elements in optical neural networks \cite{shen2017deep} with specific forms of activation functions.  Another interesting application is power regulation in photonic networks.  For example, as photonic networks for laser-driven particle accelerators \cite{hughes2018chip} must be able to handle large input powers, it may be of interest to use our approach to design compact optical limiters in these networks. For the purposes of exploring these and many other potential applications, we have made publicly available a software package that implements the algorithms discussed here \cite{hughes2018fdfdpy}.

To summarize this paper, we have developed an adjoint method, which enables gradient optimization of nonlinear photonic devices. Our work broadens the capability of inverse design for producing novel nonlinear devices. 

This work is supported by the Gordon and Betty Moore Foundation (GBMF4744); the Swiss National Science Foundation (P300P2\_177721); and the Air Force Office of Scientific Research (AFOSR) (FA9550-17-1-0002).


%

\onecolumngrid
\pagebreak
\widetext
\newpage 
\begin{center}
\section*{\textbf{\large Supplementary Information}}
\end{center}
\setcounter{equation}{0}
\setcounter{figure}{0}
\setcounter{table}{0}
\setcounter{page}{1}
\makeatletter
\renewcommand{\thepage}{S\arabic{page}}
\renewcommand{\thetable}{S\arabic{table}}
\renewcommand{\theequation}{S\arabic{equation}}
\renewcommand{\thefigure}{S\arabic{figure}}

\section{Optimization Details}

\begin{figure}[h]
\centering
\includegraphics[width=0.8\textwidth]{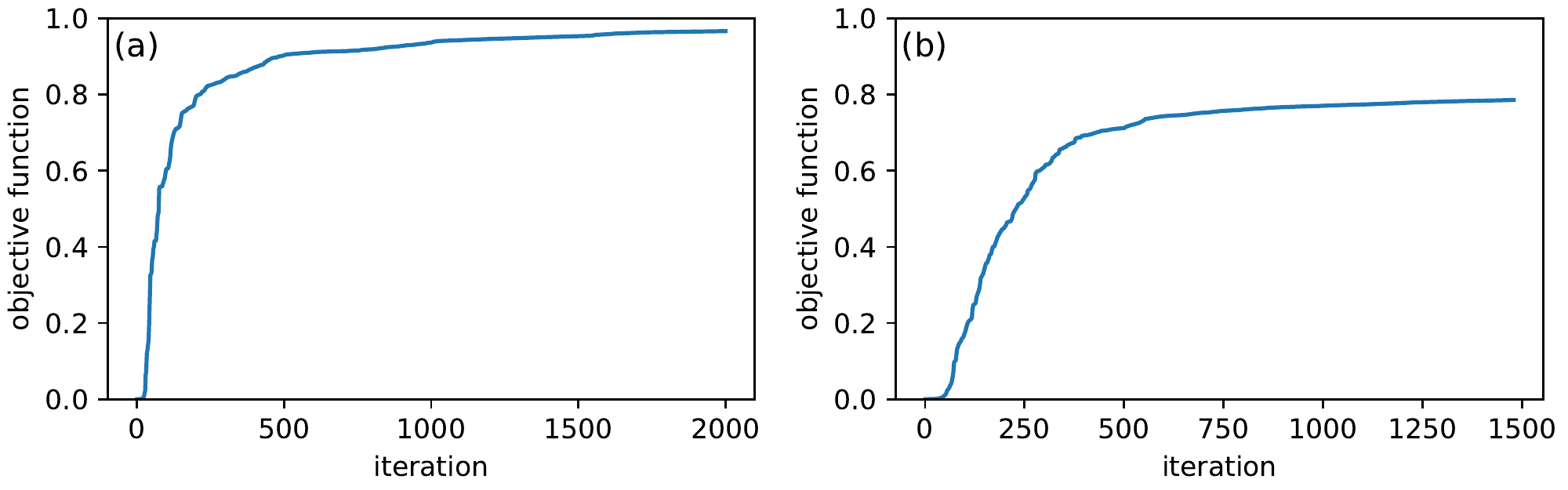}
\caption{\label{fig:objfns} Objective function vs. iteration of the optimization for (a) 2-port device of Fig. 2 and (b) 3-port device of Fig. 3 of the main text.}
\end{figure}

Table \ref{tb:params} contains the parameters used in the inverse design demonstrations of Fig. 2 and Fig. 3 of the main text. The values of the objective function vs. iteration are shown in Fig. \ref{fig:objfns} for both the 2-port and 3-port devices.

\begin{center}
\begin{table}[h]
\begin{tabular}{ |c|c|c|c|c| } 
 \hline
  parameter & symbol & value (2-port) & value (3-port) & units \\ 
  \hline
  \hline
 max relative permittivity & $\epsilon_\textrm{m}$ & 5.95 & 5.95 & - \\ 
 nonlinear susceptibility &  $\chi^{(3)}$ & $4.1\times 10^{-19}$ & $4.1\times 10^{-19}$ & m$^2$ V$^{-2}$ \\ 
 input power & $P_0$ & 157 & 57 & mW/$\mu$m \\
 free space wavelength & $\lambda_0$ & 2 & 2 & $\mu$m \\ 
 design region length & $L$ & 10 & 6 & $\mu \textrm{m}$ \\ 
 design region height & $H$ & 1.6 & 6 & $\mu \textrm{m}$ \\ 
 waveguide width & $w$ & 300 & 300 & $ \textrm{nm}$ \\ 
 grid size & g & 40 & 40 & nm \\
 low-pass filter feature size & R & 160 & 200 & nm \\
 projection strength & $\beta$ & 100 & 500 & - \\
 projection mid-point & $\eta$ & 0.5 & 0.5 & - \\
 \hline
\end{tabular}
\caption{\label{table:params} Parameters used in the optimization study.  Column `2-port' refers to the device from Fig. 2. Column `3-port' refers to the device from Fig. 3}.
\label{tb:params}
\end{table}
\end{center}

\section{Permittivity-dependent Nonlinear Susceptibility}
In the inverse design demonstration of the main text, we made the assumption that the nonlinear susceptibility distribution was proportional to the density of material in the design region.  Here we will derive the form of the adjoint sensitivity with this modification.

With our assumption, the nonlinear susceptibility vector $\bm{\chi}$ may be written in terms of the scalar magnitude of the nonlinear susceptibility $\chi^{(3)}$ and the relative permittivity vector $\bm{\epsilon}_r$ explicitly as
\begin{equation}
\bm{\chi} = 3\omega_0^2\epsilon_0 \chi^{(3)}\frac{\bm{\epsilon}_r - \mathbf{1}}{\epsilon_m - 1}
\label{eqn:X3_density}
\end{equation}
where $\mathbf{1}$ is a vector of all ones and $\epsilon_m$ is the maximum relative permittivity allowed in the optimization, corresponding to the material relative permittivity.

When choosing to relative permittivity distribution as the set of design variables, $\bm{\varphi} = \bm{\epsilon}_r$, the nonlinear adjoint problem requires the calculation of the partial derivatives $\partial \mathbf{f} / \partial \mathbf{e}$, $\partial \mathbf{f}^* / \partial \mathbf{e}$, and $\partial \mathbf{f} / \partial \bm{\epsilon}_r$.  When choosing the form of $\bm{\chi}$ from eq. (\ref{eqn:X3_density}), the partial derivatives $\partial \mathbf{f} / \partial \mathbf{e}$ and $\partial \mathbf{f}^* / \partial \mathbf{e}$ are the same as derived in the main text.  However, the term $\partial \mathbf{f} / \partial \bm{\epsilon}_r$ takes on a more complicated form given by
\begin{align}
\frac{\partial \mathbf{f}}{\partial \bm{\epsilon}_r} &= \frac{\partial}{\partial \bm{\epsilon}_r} \left[ A(\bm{\epsilon}_r)\mathbf{e} - 3\omega_0^2\epsilon_0 \chi^{(3)}\frac{\bm{\epsilon}_r - \mathbf{1}}{\epsilon_m - 1} \odot \mathbf{e} \odot \mathbf{e} \odot \mathbf{e}^* - b \right] \\ 
&= \frac{\partial A}{\partial \bm{\epsilon}_r}\mathbf{e}  - 3\omega_0^2\epsilon_0 \chi^{(3)}\frac{1}{\epsilon_m - 1} \textrm{diag}\left( \mathbf{e} \odot |\mathbf{e}|^2 \right) \\
&= \textrm{diag}\left( \mathbf{e} \right)  - 3\omega_0^2\epsilon_0 \chi^{(3)}\frac{1}{\epsilon_m - 1} \textrm{diag}\left( \mathbf{e} \odot |\mathbf{e}|^2 \right),
\label{eqn:prop_dfdp}
\end{align}
where we make use of the fact that $\left(\frac{\partial A}{\partial \bm{\epsilon}_r}\right)_{ijk} = \delta_{ij}\delta_{jk}$, where $\delta$ is the Kronecker delta.

Thus, while the adjoint field $\mathbf{e}_{aj}$ will have the same form as in the main text, when computing the sensitivity as in eq. (8) in the main text, one must insert the form of $\partial \mathbf{f} / \partial \bm{\epsilon}_r$ from eq. (\ref{eqn:prop_dfdp}).  This is in contrast to the usual case where the nonlinear susceptibility is fixed, $\partial \mathbf{f} / \partial \bm{\epsilon}_r = \textrm{diag}\left( \mathbf{e} \right)$.

\section{Maintaining minimum feature size and binarization}
To create realistic devices with sufficiently large minimum feature sizes and binarized permittivity distributions, we employed filtering and projection schemes during our optimization.  These schemes are discussed in great detail in \cite{zhou2015minimum} and related works.  

Rather than updating the permittivity distribution directly, one may instead choose to update a design density $\bm{\rho}$, which varies between 0 and 1 within the design region.  To create a structure with larger feature sizes, a low pass filter can be applied to $\bm{\rho}$ to created a filtered density, labelled $\tilde{\bm{\rho}}$:
\begin{equation}
\tilde{\rho}_i = \frac{\sum_{j\in \mathcal{D}} W_{ij} \rho_j}{\sum_{j\in \mathcal{D}} W_{ij}},
\end{equation}
where $\mathcal{D}$ denotes the design region, and $W$ is the spatial filter, defined for a feature size of $R$ as 
\begin{equation}
  W_{ij} = 
  \begin{cases}
    R - | r_{i} - r_{j} | & \text{if }  | r_{i} - r_{j} | \leq R \\
    0, & \text{otherwise } 
  \end{cases}
\end{equation}
with $|r_{i} - r_{j}|$ being the distance between points $i$ and $j$.  This defines a low-pass spatial filter on $\bm{\rho}$ with the effect of smoothing out features with length scale below $R$.

Now, for binarization of the structure, a projection scheme is used to recreate the final permittivity from the filtered density.  For this we define $\bar{\bm{\rho}}$ as the projected density, which is created from $\tilde{\bm{\rho}}$ as
\begin{equation}
\bar{\rho}_i = \frac{\tanh{\left( \beta \eta \right)} + \tanh{\left( \beta \left[ \tilde{\rho}_i - \eta \right] \right)}}{\tanh{\left( \beta \eta  \right)} + \tanh{\left( \beta \left[ 1 - \eta \right] \right)}}.
\end{equation}
Here, $\eta$ is a parameter between 0 and 1 that controls the mid-point of the projection, typically 0.5, and $\beta$ controls the strength of the projection, typically around 100.


The relative permittivity can then be determined from $\bm{\bar{\rho}}$ as
\begin{equation}
\bm{\epsilon}_r = (\epsilon_m - 1)\bm{\bar{\rho}} + 1,
\end{equation}
where $\epsilon_m$ is the maximum permittivity. 

\begin{figure}[t]
\centering
\includegraphics[width=0.8\textwidth]{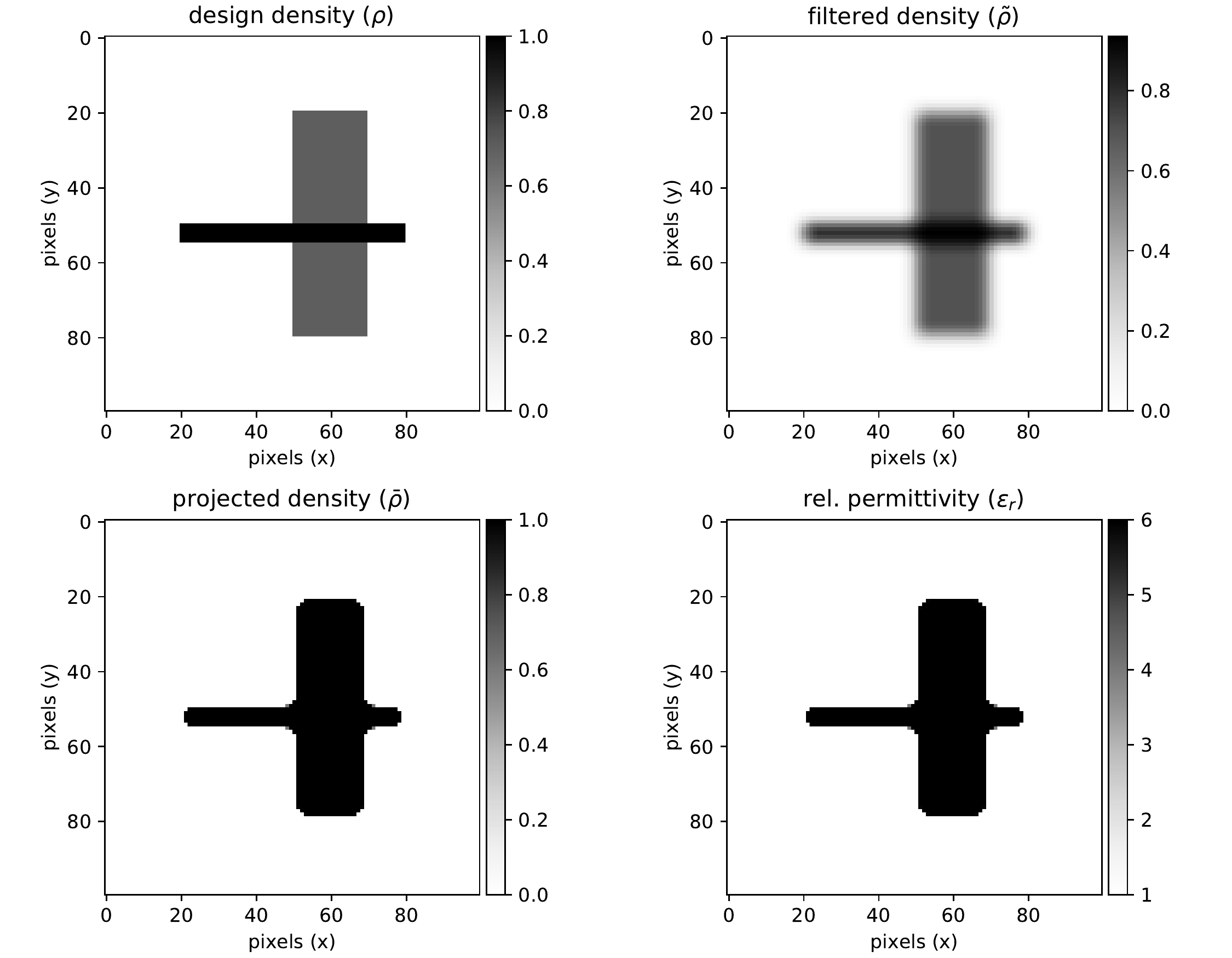}
\caption{\label{fig:filter} Filtering and projection of an example design density, $\rho$. (top left) the original density $\rho$ before processing.  (top right) the density after applying a low pass filter, $\tilde{\rho}$.  (bottom left) the density $\bar{\rho}$ after applying projection.  (bottom right) the final relative permittivity distribution $\epsilon_r$. The parameters used are $R = 200$nm, $\beta = 100$, $\eta = 0.5$.}
\end{figure}

The effect of these filtering and the binarization techniques on a sample permittivity set is illustrated in Fig. \ref{fig:filter}. In the optimizations of the main text, these techniques were performed only within the design region and required minimal modifications to the adjoint sensitivity.  The determination of $\partial \bm{\epsilon}_r / \partial \bm{\bar{\rho}}$, $\partial \bm{\bar{\rho}} / \partial \bm{\tilde{\rho}}$, and $\partial \bm{\tilde{\rho}} / \partial \bm{\rho}$ were required to compute the derivatives of the objective function with respect to the underlying $\bm{\rho}$.  For more details, see the software package accompanying this work \cite{hughes2018fdfdpy}.

\section{Nonlinear Index Shift}
Here we estimate the maximum nonlinear index shift of chalcogenide (Al$_2$S$_3$) materials.  Based on \cite{Boyd__2008, Lamont_OptExpress_2008, White_OptLett_2011}, Al$_2$S$_3$ has a nonlinear index ($n_2$) between $3 \times 10^{-14}$ and $2 \times 10^{-13}$ cm$^2$/W.
From \cite{chorel2018robust}, the damage threshold is 2.5 J/cm$^2$.  At a pulse duration of 100 ps, this damage threshold corresponds to $2.5 \times 10^{10}$ W/cm$^2$.  Together with the nonlinear refractive index, the maximum refractive index shift sustainable by the material is approximately
\begin{equation}
\Delta n = n_2~I_{damage} \approx 7.5\times10^{-4} - 5\times10^{-3}
\end{equation}
with a corresponding pulse bandwidth (for a bandwidth-limited Gaussian pulse) of $\approx 4.5$ GHz.  Our final structures exhibit maximum refractive index shifts below $5\times10^{-3}$ and their objective functions have FWHM bandwidths above 10 GHz.  This suggests that they should exhibit their desired switching effects without damage using pulse durations on the order of 100 ps and input powers on the order of 100 mW/$\mu$m.

\section{Transmission Spectra}

\begin{figure}[h]
\centering
\includegraphics[width=0.6\textwidth]{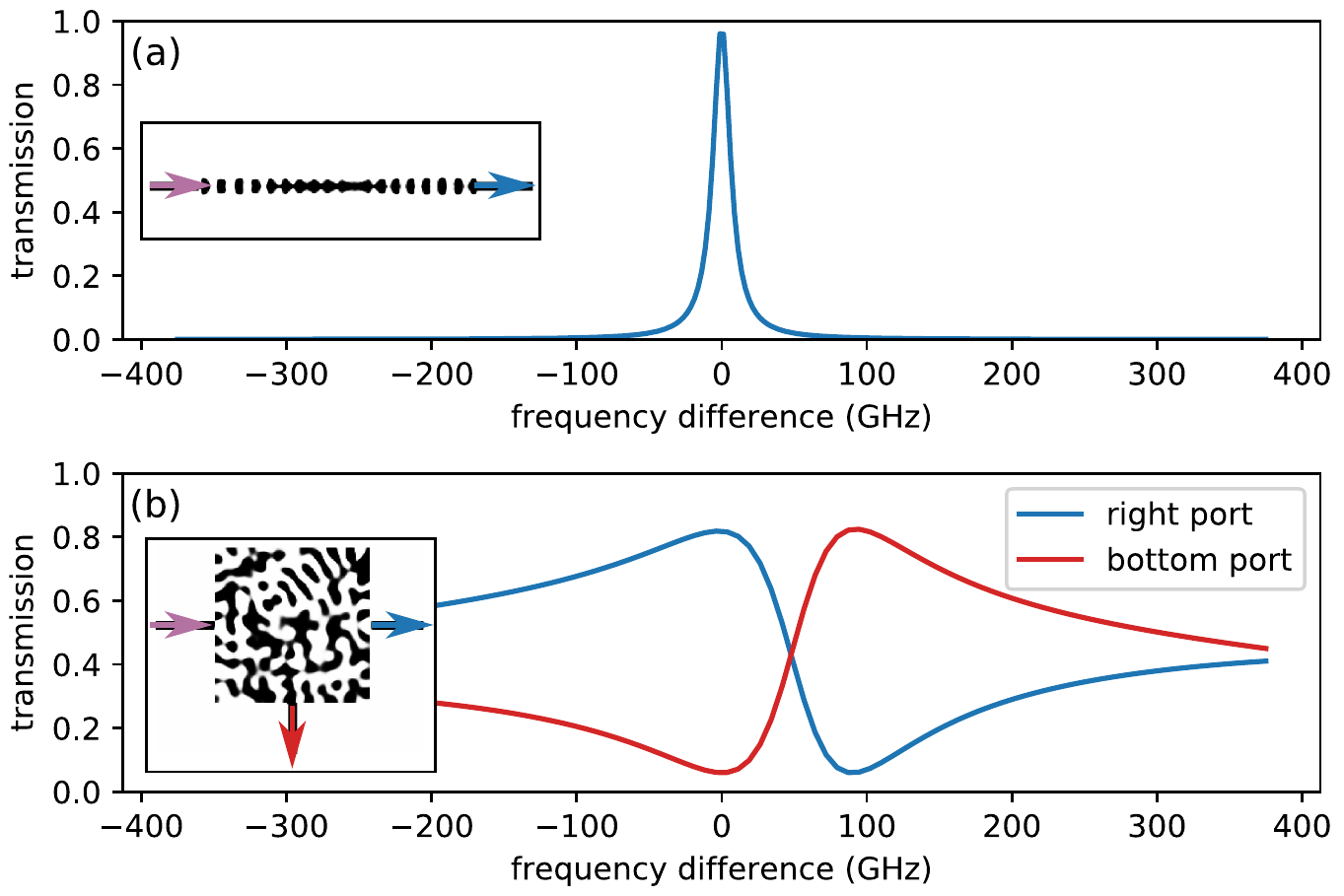}
\caption{\label{fig:spectra} (a) Transmission spectrum through the 2-port device for the low power (linear) regime. (b) Transmission spectrum through the right (blue) and bottom (red) ports of the 3-port device in the low power (linear) regime.  The x-axis represents the difference in frequency with respect to the design frequency.}
\end{figure}

The transmission vs. frequency for the two devices in the main text is shown in Fig. \ref{fig:spectra} in the linear regime.

\begin{thebibliography}{43}%
\makeatletter
\providecommand \@ifxundefined [1]{%
 \@ifx{#1\undefined}
}%
\providecommand \@ifnum [1]{%
 \ifnum #1\expandafter \@firstoftwo
 \else \expandafter \@secondoftwo
 \fi
}%
\providecommand \@ifx [1]{%
 \ifx #1\expandafter \@firstoftwo
 \else \expandafter \@secondoftwo
 \fi
}%
\providecommand \natexlab [1]{#1}%
\providecommand \enquote  [1]{``#1''}%
\providecommand \bibnamefont  [1]{#1}%
\providecommand \bibfnamefont [1]{#1}%
\providecommand \citenamefont [1]{#1}%
\providecommand \href@noop [0]{\@secondoftwo}%
\providecommand \href [0]{\begingroup \@sanitize@url \@href}%
\providecommand \@href[1]{\@@startlink{#1}\@@href}%
\providecommand \@@href[1]{\endgroup#1\@@endlink}%
\providecommand \@sanitize@url [0]{\catcode `\\12\catcode `\$12\catcode
  `\&12\catcode `\#12\catcode `\^12\catcode `\_12\catcode `\%12\relax}%
\providecommand \@@startlink[1]{}%
\providecommand \@@endlink[0]{}%
\providecommand \url  [0]{\begingroup\@sanitize@url \@url }%
\providecommand \@url [1]{\endgroup\@href {#1}{\urlprefix }}%
\providecommand \urlprefix  [0]{URL }%
\providecommand \Eprint [0]{\href }%
\providecommand \doibase [0]{http://dx.doi.org/}%
\providecommand \bibinfo  [0]{\@secondoftwo}%
\providecommand \bibfield  [0]{\@secondoftwo}%
\providecommand \translation [1]{[#1]}%
\providecommand \BibitemOpen [0]{}%
\providecommand \bibitemStop [0]{}%
\providecommand \bibitemNoStop [0]{.\EOS\space}%
\providecommand \EOS [0]{\spacefactor3000\relax}%
\providecommand \BibitemShut  [1]{\csname bibitem#1\endcsname}%
\let\auto@bib@innerbib\@empty
\bibitem [{\citenamefont {Lalau-Keraly}\ \emph {et~al.}(2013)\citenamefont
  {Lalau-Keraly}, \citenamefont {Bhargava}, \citenamefont {Miller},\ and\
  \citenamefont {Yablonovitch}}]{lalau-keraly_adjoint_2013}%
  \BibitemOpen
  \bibfield  {author} {\bibinfo {author} {\bibfnamefont {Christopher~M.}\
  \bibnamefont {Lalau-Keraly}}, \bibinfo {author} {\bibfnamefont {Samarth}\
  \bibnamefont {Bhargava}}, \bibinfo {author} {\bibfnamefont {Owen~D.}\
  \bibnamefont {Miller}}, \ and\ \bibinfo {author} {\bibfnamefont {Eli}\
  \bibnamefont {Yablonovitch}},\ }\bibfield  {title} {{
  \enquote {\bibinfo {title} {Adjoint shape optimization applied to
  electromagnetic design},}\ }}\href {\doibase 10.1364/OE.21.021693} {\bibfield
   {journal} {\bibinfo  {journal} {Optics Express}\ }\textbf {\bibinfo {volume}
  {21}},\ \bibinfo {pages} {21693--21701} (\bibinfo {year} {2013})}\BibitemShut
  {NoStop}%
\bibitem [{\citenamefont {Wang}\ \emph {et~al.}(2018)\citenamefont {Wang},
  \citenamefont {Shi}, \citenamefont {Hughes}, \citenamefont {Zhao},\ and\
  \citenamefont {Fan}}]{wang_adjoint-based_2018}%
  \BibitemOpen
  \bibfield  {author} {\bibinfo {author} {\bibfnamefont {Jiahui}\ \bibnamefont
  {Wang}}, \bibinfo {author} {\bibfnamefont {Yu}~\bibnamefont {Shi}}, \bibinfo
  {author} {\bibfnamefont {Tyler}\ \bibnamefont {Hughes}}, \bibinfo {author}
  {\bibfnamefont {Zhexin}\ \bibnamefont {Zhao}}, \ and\ \bibinfo {author}
  {\bibfnamefont {Shanhui}\ \bibnamefont {Fan}},\ }\bibfield  {title}
  {{\enquote {\bibinfo {title} {Adjoint-based optimization
  of active nanophotonic devices},}\ }}\href {\doibase 10.1364/OE.26.003236}
  {\bibfield  {journal} {\bibinfo  {journal} {Optics Express}\ }\textbf
  {\bibinfo {volume} {26}},\ \bibinfo {pages} {3236--3248} (\bibinfo {year}
  {2018})}\BibitemShut {NoStop}%
\bibitem [{\citenamefont {Elesin}\ \emph {et~al.}(2012)\citenamefont {Elesin},
  \citenamefont {Lazarov}, \citenamefont {Jensen},\ and\ \citenamefont
  {Sigmund}}]{elesin_design_2012}%
  \BibitemOpen
  \bibfield  {author} {\bibinfo {author} {\bibfnamefont {Y.}~\bibnamefont
  {Elesin}}, \bibinfo {author} {\bibfnamefont {B.~S.}\ \bibnamefont {Lazarov}},
  \bibinfo {author} {\bibfnamefont {J.~S.}\ \bibnamefont {Jensen}}, \ and\
  \bibinfo {author} {\bibfnamefont {O.}~\bibnamefont {Sigmund}},\ }\bibfield
  {title} {\enquote {\bibinfo {title} {Design of robust and efficient photonic
  switches using topology optimization},}\ }\href {\doibase
  10.1016/j.photonics.2011.10.003} {\bibfield  {journal} {\bibinfo  {journal}
  {Photonics and Nanostructures - Fundamentals and Applications}\ }\textbf
  {\bibinfo {volume} {10}},\ \bibinfo {pages} {153--165} (\bibinfo {year}
  {2012})}\BibitemShut {NoStop}%
\bibitem [{\citenamefont {Piggott}\ \emph {et~al.}(2017)\citenamefont
  {Piggott}, \citenamefont {Petykiewicz}, \citenamefont {Su},\ and\
  \citenamefont {Vu\v{c}kovi\'{c}}}]{piggott_fabrication-constrained_2017}%
  \BibitemOpen
  \bibfield  {author} {\bibinfo {author} {\bibfnamefont {Alexander~Y.}\
  \bibnamefont {Piggott}}, \bibinfo {author} {\bibfnamefont {Jan}\ \bibnamefont
  {Petykiewicz}}, \bibinfo {author} {\bibfnamefont {Logan}\ \bibnamefont {Su}},
  \ and\ \bibinfo {author} {\bibfnamefont {Jelena}\ \bibnamefont
  {Vu\v{c}kovi\'{c}}},\ }\bibfield  {title} {{\enquote
  {\bibinfo {title} {Fabrication-constrained nanophotonic inverse design},}\
  }}\href {\doibase 10.1038/s41598-017-01939-2} {\bibfield  {journal} {\bibinfo
   {journal} {Scientific Reports}\ }\textbf {\bibinfo {volume} {7}},\ \bibinfo
  {pages} {1786} (\bibinfo {year} {2017})}\BibitemShut {NoStop}%
\bibitem [{\citenamefont {Piggott}\ \emph {et~al.}(2015)\citenamefont
  {Piggott}, \citenamefont {Lu}, \citenamefont {Lagoudakis}, \citenamefont
  {Petykiewicz}, \citenamefont {Babinec},\ and\ \citenamefont
  {Vu\v{c}kovi\'{c}}}]{piggott_inverse_2015}%
  \BibitemOpen
  \bibfield  {author} {\bibinfo {author} {\bibfnamefont {Alexander~Y.}\
  \bibnamefont {Piggott}}, \bibinfo {author} {\bibfnamefont {Jesse}\
  \bibnamefont {Lu}}, \bibinfo {author} {\bibfnamefont {Konstantinos~G.}\
  \bibnamefont {Lagoudakis}}, \bibinfo {author} {\bibfnamefont {Jan}\
  \bibnamefont {Petykiewicz}}, \bibinfo {author} {\bibfnamefont {Thomas~M.}\
  \bibnamefont {Babinec}}, \ and\ \bibinfo {author} {\bibfnamefont {Jelena}\
  \bibnamefont {Vu\v{c}kovi\'{c}}},\ }\bibfield  {title} {{
  \enquote {\bibinfo {title} {Inverse design and demonstration of a compact
  and broadband on-chip wavelength demultiplexer},}\ }}\href {\doibase
  10.1038/nphoton.2015.69} {\bibfield  {journal} {\bibinfo  {journal} {Nature
  Photonics}\ }\textbf {\bibinfo {volume} {9}},\ \bibinfo {pages} {374--377}
  (\bibinfo {year} {2015})}\BibitemShut {NoStop}%
\bibitem [{\citenamefont {Kao}\ \emph {et~al.}(2005)\citenamefont {Kao},
  \citenamefont {Osher},\ and\ \citenamefont
  {Yablonovitch}}]{kao_maximizing_2005}%
  \BibitemOpen
  \bibfield  {author} {\bibinfo {author} {\bibfnamefont {C.~Y.}\ \bibnamefont
  {Kao}}, \bibinfo {author} {\bibfnamefont {S.}~\bibnamefont {Osher}}, \ and\
  \bibinfo {author} {\bibfnamefont {E.}~\bibnamefont {Yablonovitch}},\
  }\bibfield  {title} {{\enquote {\bibinfo {title}
  {Maximizing band gaps in two-dimensional photonic crystals by using level set
  methods},}\ }}\href {\doibase 10.1007/s00340-005-1877-3} {\bibfield
  {journal} {\bibinfo  {journal} {Applied Physics B}\ }\textbf {\bibinfo
  {volume} {81}},\ \bibinfo {pages} {235--244} (\bibinfo {year}
  {2005})}\BibitemShut {NoStop}%
\bibitem [{\citenamefont {Hughes}\ \emph {et~al.}(2017)\citenamefont {Hughes},
  \citenamefont {Veronis}, \citenamefont {Wootton}, \citenamefont {England},\
  and\ \citenamefont {Fan}}]{hughes_method_2017}%
  \BibitemOpen
  \bibfield  {author} {\bibinfo {author} {\bibfnamefont {Tyler}\ \bibnamefont
  {Hughes}}, \bibinfo {author} {\bibfnamefont {Georgios}\ \bibnamefont
  {Veronis}}, \bibinfo {author} {\bibfnamefont {Kent~P.}\ \bibnamefont
  {Wootton}}, \bibinfo {author} {\bibfnamefont {R.~Joel}\ \bibnamefont
  {England}}, \ and\ \bibinfo {author} {\bibfnamefont {Shanhui}\ \bibnamefont
  {Fan}},\ }\bibfield  {title} {{\enquote {\bibinfo {title}
  {Method for computationally efficient design of dielectric laser accelerator
  structures},}\ }}\href {\doibase 10.1364/OE.25.015414} {\bibfield  {journal}
  {\bibinfo  {journal} {Optics Express}\ }\textbf {\bibinfo {volume} {25}},\
  \bibinfo {pages} {15414--15427} (\bibinfo {year} {2017})}\BibitemShut
  {NoStop}%
\bibitem [{\citenamefont {Molesky}\ \emph {et~al.}(2018)\citenamefont
  {Molesky}, \citenamefont {Lin}, \citenamefont {Piggott}, \citenamefont {Jin},
  \citenamefont {Vu\v{c}kovi\'{c}},\ and\ \citenamefont
  {Rodriguez}}]{molesky_outlook_2018}%
  \BibitemOpen
  \bibfield  {author} {\bibinfo {author} {\bibfnamefont {Sean}\ \bibnamefont
  {Molesky}}, \bibinfo {author} {\bibfnamefont {Zin}\ \bibnamefont {Lin}},
  \bibinfo {author} {\bibfnamefont {Alexander~Y.}\ \bibnamefont {Piggott}},
  \bibinfo {author} {\bibfnamefont {Weiliang}\ \bibnamefont {Jin}}, \bibinfo
  {author} {\bibfnamefont {Jelena}\ \bibnamefont {Vu\v{c}kovi\'{c}}}, \ and\
  \bibinfo {author} {\bibfnamefont {Alejandro~W.}\ \bibnamefont {Rodriguez}},\
  }\bibfield  {title} {\enquote {\bibinfo {title} {Outlook for inverse design
  in nanophotonics},}\ }\href@noop {} {\bibfield  {journal} {\bibinfo
  {journal} {arXiv preprint arXiv:1801.06715}\ } (\bibinfo {year} {2018})},\
  \bibinfo {note} {arXiv: 1801.06715}\BibitemShut {NoStop}%
\bibitem [{\citenamefont {Sigmund}\ and\ \citenamefont
  {Sondergaard~Jensen}(2003)}]{sigmund_systematic_2003}%
  \BibitemOpen
  \bibfield  {author} {\bibinfo {author} {\bibfnamefont {O.}~\bibnamefont
  {Sigmund}}\ and\ \bibinfo {author} {\bibfnamefont {J.}~\bibnamefont
  {Sondergaard~Jensen}},\ }\bibfield  {title} {{\enquote
  {\bibinfo {title} {Systematic design of phononic band-gap materials and
  structures by topology optimization},}\ }}\href {\doibase
  10.1098/rsta.2003.1177} {\bibfield  {journal} {\bibinfo  {journal}
  {Philosophical Transactions of the Royal Society A: Mathematical, Physical
  and Engineering Sciences}\ }\textbf {\bibinfo {volume} {361}},\ \bibinfo
  {pages} {1001--1019} (\bibinfo {year} {2003})}\BibitemShut {NoStop}%
\bibitem [{\citenamefont {Matzen}\ \emph {et~al.}(2011)\citenamefont {Matzen},
  \citenamefont {Jensen},\ and\ \citenamefont
  {Sigmund}}]{matzen_systematic_2011}%
  \BibitemOpen
  \bibfield  {author} {\bibinfo {author} {\bibfnamefont {René}\ \bibnamefont
  {Matzen}}, \bibinfo {author} {\bibfnamefont {Jakob~S.}\ \bibnamefont
  {Jensen}}, \ and\ \bibinfo {author} {\bibfnamefont {Ole}\ \bibnamefont
  {Sigmund}},\ }\bibfield  {title} {{\enquote {\bibinfo
  {title} {Systematic design of slow-light photonic waveguides},}\ }}\href
  {\doibase 10.1364/JOSAB.28.002374} {\bibfield  {journal} {\bibinfo  {journal}
  {Journal of the Optical Society of America B}\ }\textbf {\bibinfo {volume}
  {28}},\ \bibinfo {pages} {2374--2382} (\bibinfo {year} {2011})}\BibitemShut
  {NoStop}%
\bibitem [{\citenamefont {Jensen}\ and\ \citenamefont
  {Sigmund}(2005)}]{jensen_topology_2005}%
  \BibitemOpen
  \bibfield  {author} {\bibinfo {author} {\bibfnamefont {Jakob~S.}\
  \bibnamefont {Jensen}}\ and\ \bibinfo {author} {\bibfnamefont {Ole}\
  \bibnamefont {Sigmund}},\ }\bibfield  {title} {{\enquote
  {\bibinfo {title} {Topology optimization of photonic crystal structures: a
  high-bandwidth low-loss {T}-junction waveguide},}\ }}\href {\doibase
  10.1364/JOSAB.22.001191} {\bibfield  {journal} {\bibinfo  {journal} {Journal
  of the Optical Society of America B}\ }\textbf {\bibinfo {volume} {22}},\
  \bibinfo {pages} {1191} (\bibinfo {year} {2005})}\BibitemShut {NoStop}%
\bibitem [{\citenamefont {Frellsen}\ \emph {et~al.}(2016)\citenamefont
  {Frellsen}, \citenamefont {Ding}, \citenamefont {Sigmund},\ and\
  \citenamefont {Frandsen}}]{frellsen_topology_2016}%
  \BibitemOpen
  \bibfield  {author} {\bibinfo {author} {\bibfnamefont {Louise~F.}\
  \bibnamefont {Frellsen}}, \bibinfo {author} {\bibfnamefont {Yunhong}\
  \bibnamefont {Ding}}, \bibinfo {author} {\bibfnamefont {Ole}\ \bibnamefont
  {Sigmund}}, \ and\ \bibinfo {author} {\bibfnamefont {Lars~H.}\ \bibnamefont
  {Frandsen}},\ }\bibfield  {title} {{\enquote {\bibinfo
  {title} {Topology optimized mode multiplexing in silicon-on-insulator
  photonic wire waveguides},}\ }}\href {\doibase 10.1364/OE.24.016866}
  {\bibfield  {journal} {\bibinfo  {journal} {Optics Express}\ }\textbf
  {\bibinfo {volume} {24}},\ \bibinfo {pages} {16866--16873} (\bibinfo {year}
  {2016})}\BibitemShut {NoStop}%
\bibitem [{\citenamefont {Shen}\ \emph {et~al.}(2015)\citenamefont {Shen},
  \citenamefont {Wang}, \citenamefont {Polson},\ and\ \citenamefont
  {Menon}}]{shen_integrated-nanophotonics_2015}%
  \BibitemOpen
  \bibfield  {author} {\bibinfo {author} {\bibfnamefont {Bing}\ \bibnamefont
  {Shen}}, \bibinfo {author} {\bibfnamefont {Peng}\ \bibnamefont {Wang}},
  \bibinfo {author} {\bibfnamefont {Randy}\ \bibnamefont {Polson}}, \ and\
  \bibinfo {author} {\bibfnamefont {Rajesh}\ \bibnamefont {Menon}},\ }\bibfield
   {title} {{\enquote {\bibinfo {title} {An
  integrated-nanophotonics polarization beamsplitter with 2.4 $\times$ 2.4
  $\mu$m$^{\textrm{2}}$ footprint},}\ }}\href {\doibase
  10.1038/nphoton.2015.80} {\bibfield  {journal} {\bibinfo  {journal} {Nature
  Photonics}\ }\textbf {\bibinfo {volume} {9}},\ \bibinfo {pages} {378--382}
  (\bibinfo {year} {2015})}\BibitemShut {NoStop}%
\bibitem [{\citenamefont {Shen}\ \emph {et~al.}(2003)\citenamefont {Shen},
  \citenamefont {Ye},\ and\ \citenamefont {He}}]{shen_design_2003}%
  \BibitemOpen
  \bibfield  {author} {\bibinfo {author} {\bibfnamefont {Linfang}\ \bibnamefont
  {Shen}}, \bibinfo {author} {\bibfnamefont {Zhuo}\ \bibnamefont {Ye}}, \ and\
  \bibinfo {author} {\bibfnamefont {Sailing}\ \bibnamefont {He}},\ }\bibfield
  {title} {\enquote {\bibinfo {title} {Design of two-dimensional photonic
  crystals with large absolute band gaps using a genetic algorithm},}\ }\href
  {\doibase 10.1103/PhysRevB.68.035109} {\bibfield  {journal} {\bibinfo
  {journal} {Physical Review B}\ }\textbf {\bibinfo {volume} {68}},\ \bibinfo
  {pages} {035109} (\bibinfo {year} {2003})}\BibitemShut {NoStop}%
\bibitem [{\citenamefont {Minkov}\ and\ \citenamefont
  {Savona}(2014)}]{Minkov2014}%
  \BibitemOpen
  \bibfield  {author} {\bibinfo {author} {\bibfnamefont {Momchil}\ \bibnamefont
  {Minkov}}\ and\ \bibinfo {author} {\bibfnamefont {Vincenzo}\ \bibnamefont
  {Savona}},\ }\bibfield  {title} {\enquote {\bibinfo {title} {{Automated
  optimization of photonic crystal slab cavities.}}}\ }\href {\doibase
  10.1038/srep05124} {\bibfield  {journal} {\bibinfo  {journal} {Scientific
  reports}\ }\textbf {\bibinfo {volume} {4}},\ \bibinfo {pages} {5124}
  (\bibinfo {year} {2014})}\BibitemShut {NoStop}%
\bibitem [{\citenamefont {Minkov}\ and\ \citenamefont
  {Savona}(2015)}]{Minkov2015}%
  \BibitemOpen
  \bibfield  {author} {\bibinfo {author} {\bibfnamefont {Momchil}\ \bibnamefont
  {Minkov}}\ and\ \bibinfo {author} {\bibfnamefont {Vincenzo}\ \bibnamefont
  {Savona}},\ }\bibfield  {title} {\enquote {\bibinfo {title} {Wide-band slow
  light in compact photonic crystal coupled-cavity waveguides},}\ }\href
  {\doibase 10.1364/OPTICA.2.000631} {\bibfield  {journal} {\bibinfo  {journal}
  {Optica}\ }\textbf {\bibinfo {volume} {2}},\ \bibinfo {pages} {631--634}
  (\bibinfo {year} {2015})}\BibitemShut {NoStop}%
\bibitem [{\citenamefont {Shi}\ \emph {et~al.}(2018)\citenamefont {Shi},
  \citenamefont {Li}, \citenamefont {Raman},\ and\ \citenamefont
  {Fan}}]{shi_optimization_2018}%
  \BibitemOpen
  \bibfield  {author} {\bibinfo {author} {\bibfnamefont {Yu}~\bibnamefont
  {Shi}}, \bibinfo {author} {\bibfnamefont {Wei}\ \bibnamefont {Li}}, \bibinfo
  {author} {\bibfnamefont {Aaswath}\ \bibnamefont {Raman}}, \ and\ \bibinfo
  {author} {\bibfnamefont {Shanhui}\ \bibnamefont {Fan}},\ }\bibfield  {title}
  {\enquote {\bibinfo {title} {Optimization of {Multilayer} {Optical} {Films}
  with a {Memetic} {Algorithm} and {Mixed} {Integer} {Programming}},}\ }\href
  {\doibase 10.1021/acsphotonics.7b01136} {\bibfield  {journal} {\bibinfo
  {journal} {ACS Photonics}\ }\textbf {\bibinfo {volume} {5}},\ \bibinfo
  {pages} {684--691} (\bibinfo {year} {2018})}\BibitemShut {NoStop}%
\bibitem [{\citenamefont {Veronis}\ \emph {et~al.}(2004)\citenamefont
  {Veronis}, \citenamefont {Dutton},\ and\ \citenamefont
  {Fan}}]{veronis_method_2004}%
  \BibitemOpen
  \bibfield  {author} {\bibinfo {author} {\bibfnamefont {Georgios}\
  \bibnamefont {Veronis}}, \bibinfo {author} {\bibfnamefont {Robert~W.}\
  \bibnamefont {Dutton}}, \ and\ \bibinfo {author} {\bibfnamefont {Shanhui}\
  \bibnamefont {Fan}},\ }\bibfield  {title} {{\enquote
  {\bibinfo {title} {Method for sensitivity analysis of photonic crystal
  devices},}\ }}\href {\doibase 10.1364/OL.29.002288} {\bibfield  {journal}
  {\bibinfo  {journal} {Optics Letters}\ }\textbf {\bibinfo {volume} {29}},\
  \bibinfo {pages} {2288--2290} (\bibinfo {year} {2004})}\BibitemShut {NoStop}%
\bibitem [{\citenamefont {Giles}\ and\ \citenamefont
  {Pierce}(2000)}]{giles2000introduction}%
  \BibitemOpen
  \bibfield  {author} {\bibinfo {author} {\bibfnamefont {Michael~B}\
  \bibnamefont {Giles}}\ and\ \bibinfo {author} {\bibfnamefont {Niles~A}\
  \bibnamefont {Pierce}},\ }\bibfield  {title} {\enquote {\bibinfo {title} {An
  introduction to the adjoint approach to design},}\ }\href@noop {} {\bibfield
  {journal} {\bibinfo  {journal} {Flow, turbulence and combustion}\ }\textbf
  {\bibinfo {volume} {65}},\ \bibinfo {pages} {393--415} (\bibinfo {year}
  {2000})}\BibitemShut {NoStop}%
\bibitem [{\citenamefont {Yamashita}\ \emph {et~al.}(2015)\citenamefont
  {Yamashita}, \citenamefont {Takahashi}, \citenamefont {Asano},\ and\
  \citenamefont {Noda}}]{yamashita_raman_2015}%
  \BibitemOpen
  \bibfield  {author} {\bibinfo {author} {\bibfnamefont {Daiki}\ \bibnamefont
  {Yamashita}}, \bibinfo {author} {\bibfnamefont {Yasushi}\ \bibnamefont
  {Takahashi}}, \bibinfo {author} {\bibfnamefont {Takashi}\ \bibnamefont
  {Asano}}, \ and\ \bibinfo {author} {\bibfnamefont {Susumu}\ \bibnamefont
  {Noda}},\ }\bibfield  {title} {{\enquote {\bibinfo
  {title} {Raman shift and strain effect in high-{Q} photonic crystal silicon
  nanocavity},}\ }}\href {\doibase 10.1364/OE.23.003951} {\bibfield  {journal}
  {\bibinfo  {journal} {Optics Express}\ }\textbf {\bibinfo {volume} {23}},\
  \bibinfo {pages} {3951} (\bibinfo {year} {2015})}\BibitemShut {NoStop}%
\bibitem [{\citenamefont {Okawachi}\ \emph {et~al.}(2011)\citenamefont
  {Okawachi}, \citenamefont {Saha}, \citenamefont {Levy}, \citenamefont {Wen},
  \citenamefont {Lipson},\ and\ \citenamefont
  {Gaeta}}]{okawachi_octave-spanning_2011}%
  \BibitemOpen
  \bibfield  {author} {\bibinfo {author} {\bibfnamefont {Yoshitomo}\
  \bibnamefont {Okawachi}}, \bibinfo {author} {\bibfnamefont {Kasturi}\
  \bibnamefont {Saha}}, \bibinfo {author} {\bibfnamefont {Jacob~S.}\
  \bibnamefont {Levy}}, \bibinfo {author} {\bibfnamefont {Y.~Henry}\
  \bibnamefont {Wen}}, \bibinfo {author} {\bibfnamefont {Michal}\ \bibnamefont
  {Lipson}}, \ and\ \bibinfo {author} {\bibfnamefont {Alexander~L.}\
  \bibnamefont {Gaeta}},\ }\bibfield  {title} {{\enquote
  {\bibinfo {title} {Octave-spanning frequency comb generation in a silicon
  nitride chip},}\ }}\href {\doibase 10.1364/OL.36.003398} {\bibfield
  {journal} {\bibinfo  {journal} {Optics Letters}\ }\textbf {\bibinfo {volume}
  {36}},\ \bibinfo {pages} {3398} (\bibinfo {year} {2011})}\BibitemShut
  {NoStop}%
\bibitem [{\citenamefont {Moon}\ \emph {et~al.}(1997)\citenamefont {Moon},
  \citenamefont {Kim}, \citenamefont {Kim}, \citenamefont {Kang}, \citenamefont
  {Kim}, \citenamefont {Kim}, \citenamefont {Hahn},\ and\ \citenamefont
  {Park}}]{moon_absolute_1997}%
  \BibitemOpen
  \bibfield  {author} {\bibinfo {author} {\bibfnamefont {Joong~Ho}\
  \bibnamefont {Moon}}, \bibinfo {author} {\bibfnamefont {Jin~Ho}\ \bibnamefont
  {Kim}}, \bibinfo {author} {\bibfnamefont {Ki-jeong}\ \bibnamefont {Kim}},
  \bibinfo {author} {\bibfnamefont {Tai-Hee}\ \bibnamefont {Kang}}, \bibinfo
  {author} {\bibfnamefont {Bongsoo}\ \bibnamefont {Kim}}, \bibinfo {author}
  {\bibfnamefont {Chan-Ho}\ \bibnamefont {Kim}}, \bibinfo {author}
  {\bibfnamefont {Jong~Hoon}\ \bibnamefont {Hahn}}, \ and\ \bibinfo {author}
  {\bibfnamefont {Joon~Won}\ \bibnamefont {Park}},\ }\bibfield  {title}
  {{\enquote {\bibinfo {title} {Absolute {Surface}
  {Density} of the {Amine} {Group} of the {Aminosilylated} {Thin} {Layers}:
  {Ultraviolet}-{Visible} {Spectroscopy}, {Second} {Harmonic} {Generation}, and
  {Synchrotron}-{Radiation} {Photoelectron} {Spectroscopy} {Study}},}\ }}\href
  {\doibase 10.1021/la9705118} {\bibfield  {journal} {\bibinfo  {journal}
  {Langmuir}\ }\textbf {\bibinfo {volume} {13}},\ \bibinfo {pages} {4305--4310}
  (\bibinfo {year} {1997})}\BibitemShut {NoStop}%
\bibitem [{\citenamefont {Khoram}\ \emph {et~al.}(2018)\citenamefont {Khoram},
  \citenamefont {Chen}, \citenamefont {Liu}, \citenamefont {Wang},\ and\
  \citenamefont {Yu}}]{khoram2018stochastic}%
  \BibitemOpen
  \bibfield  {author} {\bibinfo {author} {\bibfnamefont {Erfan}\ \bibnamefont
  {Khoram}}, \bibinfo {author} {\bibfnamefont {Ang}\ \bibnamefont {Chen}},
  \bibinfo {author} {\bibfnamefont {Dianjing}\ \bibnamefont {Liu}}, \bibinfo
  {author} {\bibfnamefont {Qiqi}\ \bibnamefont {Wang}}, \ and\ \bibinfo
  {author} {\bibfnamefont {Zongfu}\ \bibnamefont {Yu}},\ }\bibfield  {title}
  {\enquote {\bibinfo {title} {Stochastic optimization of nonlinear
  nanophotonic media for artificial neural inference},}\ }\href@noop {}
  {\bibfield  {journal} {\bibinfo  {journal} {arXiv preprint arXiv:1810.07815}\
  } (\bibinfo {year} {2018})}\BibitemShut {NoStop}%
\bibitem [{\citenamefont {Guo}\ \emph {et~al.}(2016)\citenamefont {Guo},
  \citenamefont {Zou}, \citenamefont {Jung},\ and\ \citenamefont
  {Tang}}]{guo_-chip_2016}%
  \BibitemOpen
  \bibfield  {author} {\bibinfo {author} {\bibfnamefont {Xiang}\ \bibnamefont
  {Guo}}, \bibinfo {author} {\bibfnamefont {Chang-Ling}\ \bibnamefont {Zou}},
  \bibinfo {author} {\bibfnamefont {Hojoong}\ \bibnamefont {Jung}}, \ and\
  \bibinfo {author} {\bibfnamefont {Hong~X.}\ \bibnamefont {Tang}},\ }\bibfield
   {title} {{\enquote {\bibinfo {title} {On-{Chip} {Strong}
  {Coupling} and {Efficient} {Frequency} {Conversion} between {Telecom} and
  {Visible} {Optical} {Modes}},}\ }}\href {\doibase
  10.1103/PhysRevLett.117.123902} {\bibfield  {journal} {\bibinfo  {journal}
  {Physical Review Letters}\ }\textbf {\bibinfo {volume} {117}} (\bibinfo
  {year} {2016}),\ 10.1103/PhysRevLett.117.123902}\BibitemShut {NoStop}%
\bibitem [{\citenamefont {Lin}\ \emph {et~al.}(2016)\citenamefont {Lin},
  \citenamefont {Liang}, \citenamefont {Lon\v{c}ar}, \citenamefont {Johnson},\
  and\ \citenamefont {Rodriguez}}]{lin_cavity-enhanced_2016}%
  \BibitemOpen
  \bibfield  {author} {\bibinfo {author} {\bibfnamefont {Zin}\ \bibnamefont
  {Lin}}, \bibinfo {author} {\bibfnamefont {Xiangdong}\ \bibnamefont {Liang}},
  \bibinfo {author} {\bibfnamefont {Marko}\ \bibnamefont {Lon\v{c}ar}},
  \bibinfo {author} {\bibfnamefont {Steven~G.}\ \bibnamefont {Johnson}}, \ and\
  \bibinfo {author} {\bibfnamefont {Alejandro~W.}\ \bibnamefont {Rodriguez}},\
  }\bibfield  {title} {{\enquote {\bibinfo {title}
  {Cavity-enhanced second-harmonic generation via nonlinear-overlap
  optimization},}\ }}\href {\doibase 10.1364/OPTICA.3.000233} {\bibfield
  {journal} {\bibinfo  {journal} {Optica}\ }\textbf {\bibinfo {volume} {3}},\
  \bibinfo {pages} {233} (\bibinfo {year} {2016})}\BibitemShut {NoStop}%
\bibitem [{\citenamefont {Lin}\ \emph {et~al.}(2017)\citenamefont {Lin},
  \citenamefont {Lon\v{c}ar},\ and\ \citenamefont
  {Rodriguez}}]{lin_topology_2017}%
  \BibitemOpen
  \bibfield  {author} {\bibinfo {author} {\bibfnamefont {Zin}\ \bibnamefont
  {Lin}}, \bibinfo {author} {\bibfnamefont {Marko}\ \bibnamefont {Lon\v{c}ar}},
  \ and\ \bibinfo {author} {\bibfnamefont {Alejandro~W.}\ \bibnamefont
  {Rodriguez}},\ }\bibfield  {title} {{\enquote {\bibinfo
  {title} {Topology optimization of multi-track ring resonators and 2d
  microcavities for nonlinear frequency conversion},}\ }}\href {\doibase
  10.1364/OL.42.002818} {\bibfield  {journal} {\bibinfo  {journal} {Optics
  Letters}\ }\textbf {\bibinfo {volume} {42}},\ \bibinfo {pages} {2818}
  (\bibinfo {year} {2017})}\BibitemShut {NoStop}%
\bibitem [{\citenamefont {Bravo-Abad}\ \emph {et~al.}(2007)\citenamefont
  {Bravo-Abad}, \citenamefont {Rodriguez}, \citenamefont {Bermel},
  \citenamefont {Johnson}, \citenamefont {Joannopoulos},\ and\ \citenamefont
  {Solja{\v{c}}i{\'{c}}}}]{bravo-abad_enhanced_2007}%
  \BibitemOpen
  \bibfield  {author} {\bibinfo {author} {\bibfnamefont {Jorge}\ \bibnamefont
  {Bravo-Abad}}, \bibinfo {author} {\bibfnamefont {Alejandro}\ \bibnamefont
  {Rodriguez}}, \bibinfo {author} {\bibfnamefont {Peter}\ \bibnamefont
  {Bermel}}, \bibinfo {author} {\bibfnamefont {Steven~G.}\ \bibnamefont
  {Johnson}}, \bibinfo {author} {\bibfnamefont {John~D.}\ \bibnamefont
  {Joannopoulos}}, \ and\ \bibinfo {author} {\bibfnamefont {Marin}\
  \bibnamefont {Solja{\v{c}}i{\'{c}}}},\ }\bibfield  {title} {{
  \enquote {\bibinfo {title} {Enhanced nonlinear optics in photonic-crystal
  microcavities},}\ }}\href {\doibase 10.1364/OE.15.016161} {\bibfield
  {journal} {\bibinfo  {journal} {Optics Express}\ }\textbf {\bibinfo {volume}
  {15}},\ \bibinfo {pages} {16161--16176} (\bibinfo {year} {2007})}\BibitemShut
  {NoStop}%
\bibitem [{\citenamefont {Strang}(2007)}]{Strang2007}%
  \BibitemOpen
  \bibfield  {author} {\bibinfo {author} {\bibfnamefont {Gilbert}\ \bibnamefont
  {Strang}},\ }\href@noop {} {\emph {\bibinfo {title} {Computational Science
  and Engineering}}}\ (\bibinfo  {publisher} {Wellesley-Cambridge Press},\
  \bibinfo {year} {2007})\ \bibinfo {note} {chapter 8}\BibitemShut {NoStop}%
\bibitem [{\citenamefont {Press}\ \emph {et~al.}(2007)\citenamefont {Press},
  \citenamefont {Teukolsky}, \citenamefont {Vetterling},\ and\ \citenamefont
  {Flannery}}]{press2007numerical}%
  \BibitemOpen
  \bibfield  {author} {\bibinfo {author} {\bibfnamefont {William~H}\
  \bibnamefont {Press}}, \bibinfo {author} {\bibfnamefont {Saul~A}\
  \bibnamefont {Teukolsky}}, \bibinfo {author} {\bibfnamefont {William~T}\
  \bibnamefont {Vetterling}}, \ and\ \bibinfo {author} {\bibfnamefont
  {Brian~P}\ \bibnamefont {Flannery}},\ }\href@noop {} {\emph {\bibinfo {title}
  {Numerical recipes 3rd edition: The art of scientific computing}}}\ (\bibinfo
   {publisher} {Cambridge university press},\ \bibinfo {year}
  {2007})\BibitemShut {NoStop}%
\bibitem [{\citenamefont {Boyd}(2008)}]{Boyd__2008}%
  \BibitemOpen
  \bibfield  {author} {\bibinfo {author} {\bibfnamefont {Robert~W.}\
  \bibnamefont {Boyd}},\ }\href@noop {} {{\emph {\bibinfo
  {title} {Nonlinear {{Optics}}}}}}\ (\bibinfo  {publisher} {{Academic
  Press}},\ \bibinfo {year} {2008})\BibitemShut {NoStop}%
\bibitem [{\citenamefont {Shin}\ and\ \citenamefont
  {Fan}(2012)}]{shin2012choice}%
  \BibitemOpen
  \bibfield  {author} {\bibinfo {author} {\bibfnamefont {Wonseok}\ \bibnamefont
  {Shin}}\ and\ \bibinfo {author} {\bibfnamefont {Shanhui}\ \bibnamefont
  {Fan}},\ }\bibfield  {title} {\enquote {\bibinfo {title} {Choice of the
  perfectly matched layer boundary condition for frequency-domain maxwell's
  equations solvers},}\ }\href@noop {} {\bibfield  {journal} {\bibinfo
  {journal} {Journal of Computational Physics}\ }\textbf {\bibinfo {volume}
  {231}},\ \bibinfo {pages} {3406--3431} (\bibinfo {year} {2012})}\BibitemShut
  {NoStop}%
\bibitem [{\citenamefont {Yee}(1966)}]{yee1966numerical}%
  \BibitemOpen
  \bibfield  {author} {\bibinfo {author} {\bibfnamefont {Kane}\ \bibnamefont
  {Yee}},\ }\bibfield  {title} {\enquote {\bibinfo {title} {Numerical solution
  of initial boundary value problems involving maxwell's equations in isotropic
  media},}\ }\href@noop {} {\bibfield  {journal} {\bibinfo  {journal} {IEEE
  Transactions on antennas and propagation}\ }\textbf {\bibinfo {volume}
  {14}},\ \bibinfo {pages} {302--307} (\bibinfo {year} {1966})}\BibitemShut
  {NoStop}%
\bibitem [{\citenamefont {Byrd}\ \emph {et~al.}(1995)\citenamefont {Byrd},
  \citenamefont {Lu}, \citenamefont {Nocedal},\ and\ \citenamefont
  {Zhu}}]{byrd1995limited}%
  \BibitemOpen
  \bibfield  {author} {\bibinfo {author} {\bibfnamefont {Richard~H}\
  \bibnamefont {Byrd}}, \bibinfo {author} {\bibfnamefont {Peihuang}\
  \bibnamefont {Lu}}, \bibinfo {author} {\bibfnamefont {Jorge}\ \bibnamefont
  {Nocedal}}, \ and\ \bibinfo {author} {\bibfnamefont {Ciyou}\ \bibnamefont
  {Zhu}},\ }\bibfield  {title} {\enquote {\bibinfo {title} {A limited memory
  algorithm for bound constrained optimization},}\ }\href@noop {} {\bibfield
  {journal} {\bibinfo  {journal} {SIAM Journal on Scientific Computing}\
  }\textbf {\bibinfo {volume} {16}},\ \bibinfo {pages} {1190--1208} (\bibinfo
  {year} {1995})}\BibitemShut {NoStop}%
\bibitem [{\citenamefont {White}\ and\ \citenamefont
  {Monro}(2011)}]{White_OptLett_2011}%
  \BibitemOpen
  \bibfield  {author} {\bibinfo {author} {\bibfnamefont {Richard~T.}\
  \bibnamefont {White}}\ and\ \bibinfo {author} {\bibfnamefont {Tanya~M.}\
  \bibnamefont {Monro}},\ }\bibfield  {title} {{\enquote
  {\bibinfo {title} {Cascaded {{Raman}} shifting of high-peak-power nanosecond
  pulses in as$_2$s$_3$ and as$_2$se$_3$ optical fibers},}\ }}\href {\doibase
  10.1364/OL.36.002351} {\bibfield  {journal} {\bibinfo  {journal} {Optics
  Letters}\ }\textbf {\bibinfo {volume} {36}},\ \bibinfo {pages} {2351}
  (\bibinfo {year} {2011})}\BibitemShut {NoStop}%
\bibitem [{\citenamefont {Lamont}\ \emph {et~al.}(2008)\citenamefont {Lamont},
  \citenamefont {{Luther-Davies}}, \citenamefont {Choi}, \citenamefont
  {Madden},\ and\ \citenamefont {Eggleton}}]{Lamont_OptExpress_2008}%
  \BibitemOpen
  \bibfield  {author} {\bibinfo {author} {\bibfnamefont {Michael~R.}\
  \bibnamefont {Lamont}}, \bibinfo {author} {\bibfnamefont {Barry}\
  \bibnamefont {{Luther-Davies}}}, \bibinfo {author} {\bibfnamefont {Duk-Yong}\
  \bibnamefont {Choi}}, \bibinfo {author} {\bibfnamefont {Steve}\ \bibnamefont
  {Madden}}, \ and\ \bibinfo {author} {\bibfnamefont {Benjamin~J.}\
  \bibnamefont {Eggleton}},\ }\bibfield  {title} {{\enquote
  {\bibinfo {title} {Supercontinuum generation in dispersion engineered highly
  nonlinear ($\gamma = 10$ w/m) as$_2$s$_3$ chalcogenide planar waveguide},}\
  }}\href {\doibase 10.1364/OE.16.014938} {\bibfield  {journal} {\bibinfo
  {journal} {Optics Express}\ }\textbf {\bibinfo {volume} {16}},\ \bibinfo
  {pages} {14938} (\bibinfo {year} {2008})}\BibitemShut {NoStop}%
\bibitem [{\citenamefont {Zhou}\ \emph {et~al.}(2015)\citenamefont {Zhou},
  \citenamefont {Lazarov}, \citenamefont {Wang},\ and\ \citenamefont
  {Sigmund}}]{zhou2015minimum}%
  \BibitemOpen
  \bibfield  {author} {\bibinfo {author} {\bibfnamefont {Mingdong}\
  \bibnamefont {Zhou}}, \bibinfo {author} {\bibfnamefont {Boyan~S}\
  \bibnamefont {Lazarov}}, \bibinfo {author} {\bibfnamefont {Fengwen}\
  \bibnamefont {Wang}}, \ and\ \bibinfo {author} {\bibfnamefont {Ole}\
  \bibnamefont {Sigmund}},\ }\bibfield  {title} {\enquote {\bibinfo {title}
  {Minimum length scale in topology optimization by geometric constraints},}\
  }\href@noop {} {\bibfield  {journal} {\bibinfo  {journal} {Computer Methods
  in Applied Mechanics and Engineering}\ }\textbf {\bibinfo {volume} {293}},\
  \bibinfo {pages} {266--282} (\bibinfo {year} {2015})}\BibitemShut {NoStop}%
\bibitem [{\citenamefont {Solja\ifmmode \check{c}\else
  \v{c}\fi{}i\ifmmode~\acute{c}\else \'{c}\fi{}}\ \emph
  {et~al.}(2002)\citenamefont {Solja\ifmmode \check{c}\else
  \v{c}\fi{}i\ifmmode~\acute{c}\else \'{c}\fi{}}, \citenamefont {Ibanescu},
  \citenamefont {Johnson}, \citenamefont {Fink},\ and\ \citenamefont
  {Joannopoulos}}]{soljacic2002optimal}%
  \BibitemOpen
  \bibfield  {author} {\bibinfo {author} {\bibfnamefont {Marin}\ \bibnamefont
  {Solja\ifmmode \check{c}\else \v{c}\fi{}i\ifmmode~\acute{c}\else
  \'{c}\fi{}}}, \bibinfo {author} {\bibfnamefont {Mihai}\ \bibnamefont
  {Ibanescu}}, \bibinfo {author} {\bibfnamefont {Steven~G.}\ \bibnamefont
  {Johnson}}, \bibinfo {author} {\bibfnamefont {Yoel}\ \bibnamefont {Fink}}, \
  and\ \bibinfo {author} {\bibfnamefont {J.~D.}\ \bibnamefont {Joannopoulos}},\
  }\bibfield  {title} {\enquote {\bibinfo {title} {Optimal bistable switching
  in nonlinear photonic crystals},}\ }\href {\doibase
  10.1103/PhysRevE.66.055601} {\bibfield  {journal} {\bibinfo  {journal} {Phys.
  Rev. E}\ }\textbf {\bibinfo {volume} {66}},\ \bibinfo {pages} {055601}
  (\bibinfo {year} {2002})}\BibitemShut {NoStop}%
\bibitem [{\citenamefont {Yanik}\ \emph {et~al.}(2003)\citenamefont {Yanik},
  \citenamefont {Fan}, \citenamefont {Solja{\v{c}}i{\'{c}}},\ and\
  \citenamefont {Joannopoulos}}]{yanik_all-optical_2003}%
  \BibitemOpen
  \bibfield  {author} {\bibinfo {author} {\bibfnamefont {Mehmet~Fatih}\
  \bibnamefont {Yanik}}, \bibinfo {author} {\bibfnamefont {Shanhui}\
  \bibnamefont {Fan}}, \bibinfo {author} {\bibfnamefont {Marin}\ \bibnamefont
  {Solja{\v{c}}i{\'{c}}}}, \ and\ \bibinfo {author} {\bibfnamefont {J.~D.}\
  \bibnamefont {Joannopoulos}},\ }\bibfield  {title} {{
  \enquote {\bibinfo {title} {All-optical transistor action with bistable
  switching in a photonic crystal cross-waveguide geometry},}\ }}\href
  {\doibase 10.1364/OL.28.002506} {\bibfield  {journal} {\bibinfo  {journal}
  {Optics Letters}\ }\textbf {\bibinfo {volume} {28}},\ \bibinfo {pages} {2506}
  (\bibinfo {year} {2003})}\BibitemShut {NoStop}%
\bibitem [{\citenamefont {Chorel}\ \emph {et~al.}(2018)\citenamefont {Chorel},
  \citenamefont {Lanternier}, \citenamefont {Lavastre}, \citenamefont {Bonod},
  \citenamefont {Bousquet},\ and\ \citenamefont
  {N{\'e}auport}}]{chorel2018robust}%
  \BibitemOpen
  \bibfield  {author} {\bibinfo {author} {\bibfnamefont {Marine}\ \bibnamefont
  {Chorel}}, \bibinfo {author} {\bibfnamefont {Thomas}\ \bibnamefont
  {Lanternier}}, \bibinfo {author} {\bibfnamefont {Eric}\ \bibnamefont
  {Lavastre}}, \bibinfo {author} {\bibfnamefont {Nicolas}\ \bibnamefont
  {Bonod}}, \bibinfo {author} {\bibfnamefont {Bruno}\ \bibnamefont {Bousquet}},
  \ and\ \bibinfo {author} {\bibfnamefont {J{\'e}r{\^o}me}\ \bibnamefont
  {N{\'e}auport}},\ }\bibfield  {title} {\enquote {\bibinfo {title} {Robust
  optimization of the laser induced damage threshold of dielectric mirrors for
  high power lasers},}\ }\href@noop {} {\bibfield  {journal} {\bibinfo
  {journal} {Optics express}\ }\textbf {\bibinfo {volume} {26}},\ \bibinfo
  {pages} {11764--11774} (\bibinfo {year} {2018})}\BibitemShut {NoStop}%
\bibitem [{\citenamefont {Shi}\ \emph {et~al.}(2016)\citenamefont {Shi},
  \citenamefont {Shin},\ and\ \citenamefont {Fan}}]{Shi2016multi}%
  \BibitemOpen
  \bibfield  {author} {\bibinfo {author} {\bibfnamefont {Yu}~\bibnamefont
  {Shi}}, \bibinfo {author} {\bibfnamefont {Wonseok}\ \bibnamefont {Shin}}, \
  and\ \bibinfo {author} {\bibfnamefont {Shanhui}\ \bibnamefont {Fan}},\
  }\bibfield  {title} {\enquote {\bibinfo {title} {Multi-frequency
  finite-difference frequency-domain algorithm for active nanophotonic device
  simulations},}\ }\href {\doibase 10.1364/OPTICA.3.001256} {\bibfield
  {journal} {\bibinfo  {journal} {Optica}\ }\textbf {\bibinfo {volume} {3}},\
  \bibinfo {pages} {1256--1259} (\bibinfo {year} {2016})}\BibitemShut {NoStop}%
\bibitem [{\citenamefont {Shen}\ \emph {et~al.}(2017)\citenamefont {Shen},
  \citenamefont {Harris}, \citenamefont {Skirlo}, \citenamefont {Prabhu},
  \citenamefont {Baehr-Jones}, \citenamefont {Hochberg}, \citenamefont {Sun},
  \citenamefont {Zhao}, \citenamefont {Larochelle}, \citenamefont {Englund},\
  and\ \citenamefont {Solja{\v{c}}i{\'{c}}}}]{shen2017deep}%
  \BibitemOpen
  \bibfield  {author} {\bibinfo {author} {\bibfnamefont {Yichen}\ \bibnamefont
  {Shen}}, \bibinfo {author} {\bibfnamefont {Nicholas~C}\ \bibnamefont
  {Harris}}, \bibinfo {author} {\bibfnamefont {Scott}\ \bibnamefont {Skirlo}},
  \bibinfo {author} {\bibfnamefont {Mihika}\ \bibnamefont {Prabhu}}, \bibinfo
  {author} {\bibfnamefont {Tom}\ \bibnamefont {Baehr-Jones}}, \bibinfo {author}
  {\bibfnamefont {Michael}\ \bibnamefont {Hochberg}}, \bibinfo {author}
  {\bibfnamefont {Xin}\ \bibnamefont {Sun}}, \bibinfo {author} {\bibfnamefont
  {Shijie}\ \bibnamefont {Zhao}}, \bibinfo {author} {\bibfnamefont {Hugo}\
  \bibnamefont {Larochelle}}, \bibinfo {author} {\bibfnamefont {Dirk}\
  \bibnamefont {Englund}}, \ and\ \bibinfo {author} {\bibfnamefont {Marin}\
  \bibnamefont {Solja{\v{c}}i{\'{c}}}},\ }\bibfield  {title} {\enquote
  {\bibinfo {title} {Deep learning with coherent nanophotonic circuits},}\
  }\href@noop {} {\bibfield  {journal} {\bibinfo  {journal} {Nature Photonics}\
  }\textbf {\bibinfo {volume} {11}},\ \bibinfo {pages} {441} (\bibinfo {year}
  {2017})}\BibitemShut {NoStop}%
\bibitem [{\citenamefont {Hughes}\ \emph
  {et~al.}(2018{\natexlab{a}})\citenamefont {Hughes}, \citenamefont {Tan},
  \citenamefont {Zhao}, \citenamefont {Sapra}, \citenamefont {Leedle},
  \citenamefont {Deng}, \citenamefont {Miao}, \citenamefont {Black},
  \citenamefont {Solgaard}, \citenamefont {Harris}, \citenamefont {Vuckovic},
  \citenamefont {Byer}, \citenamefont {Fan}, \citenamefont {England},
  \citenamefont {Lee},\ and\ \citenamefont {Qi}}]{hughes2018chip}%
  \BibitemOpen
  \bibfield  {author} {\bibinfo {author} {\bibfnamefont {Tyler~W.}\
  \bibnamefont {Hughes}}, \bibinfo {author} {\bibfnamefont {Si}~\bibnamefont
  {Tan}}, \bibinfo {author} {\bibfnamefont {Zhexin}\ \bibnamefont {Zhao}},
  \bibinfo {author} {\bibfnamefont {Neil~V.}\ \bibnamefont {Sapra}}, \bibinfo
  {author} {\bibfnamefont {Kenneth~J.}\ \bibnamefont {Leedle}}, \bibinfo
  {author} {\bibfnamefont {Huiyang}\ \bibnamefont {Deng}}, \bibinfo {author}
  {\bibfnamefont {Yu}~\bibnamefont {Miao}}, \bibinfo {author} {\bibfnamefont
  {Dylan~S.}\ \bibnamefont {Black}}, \bibinfo {author} {\bibfnamefont {Olav}\
  \bibnamefont {Solgaard}}, \bibinfo {author} {\bibfnamefont {James~S.}\
  \bibnamefont {Harris}}, \bibinfo {author} {\bibfnamefont {Jelena}\
  \bibnamefont {Vuckovic}}, \bibinfo {author} {\bibfnamefont {Robert~L.}\
  \bibnamefont {Byer}}, \bibinfo {author} {\bibfnamefont {Shanhui}\
  \bibnamefont {Fan}}, \bibinfo {author} {\bibfnamefont {R.~Joel}\ \bibnamefont
  {England}}, \bibinfo {author} {\bibfnamefont {Yun~Jo}\ \bibnamefont {Lee}}, \
  and\ \bibinfo {author} {\bibfnamefont {Minghao}\ \bibnamefont {Qi}},\
  }\bibfield  {title} {\enquote {\bibinfo {title} {On-chip laser-power delivery
  system for dielectric laser accelerators},}\ }\href@noop {} {\bibfield
  {journal} {\bibinfo  {journal} {Physical Review Applied}\ }\textbf {\bibinfo
  {volume} {9}},\ \bibinfo {pages} {054017} (\bibinfo {year}
  {2018}{\natexlab{a}})}\BibitemShut {NoStop}%
\bibitem [{\citenamefont {Hughes}\ \emph
  {et~al.}(2018{\natexlab{b}})\citenamefont {Hughes}, \citenamefont {Minkov},\
  and\ \citenamefont {Williamson}}]{hughes2018fdfdpy}%
  \BibitemOpen
  \bibfield  {author} {\bibinfo {author} {\bibfnamefont {Tyler~W}\ \bibnamefont
  {Hughes}}, \bibinfo {author} {\bibfnamefont {Momchil}\ \bibnamefont
  {Minkov}}, \ and\ \bibinfo {author} {\bibfnamefont {Ian A~D}\ \bibnamefont
  {Williamson}},\ }\href@noop {} {\enquote {\bibinfo {title} {{\em Angler} --
  an adjoint nonliner gradient open-source package},}\ }\bibinfo {howpublished}
  {\url{https://github.com/fancompute/angler}} (\bibinfo {year}
  {2018}{\natexlab{b}})\BibitemShut {NoStop}%
\end{thebibliography}
\end{document}